\documentclass[useAMS,usenatbib,fleqn]{mn2e}       
\usepackage{graphicx}
\usepackage{amssymb}
\usepackage{amsmath}
\usepackage{aas_macros}
\usepackage{deluxetable}
\usepackage{lscape}
\usepackage{subfigure}
\usepackage{hyperref}

\begin{document}

\title[Kompaneets Fitting of Ori--Eri Superbubble]{Kompaneets Model Fitting of the Orion--Eridanus Superbubble}

\author[A. Pon et al.] {Andy Pon, $^{1,2,3,4}$ Doug Johnstone, $^{5,4,3}$ John Bally $^{6}$ \& Carl Heiles $^7$ \\
$^1$Max Planck Institute for Extraterrestrial Physics, Giessenbachstrasse 1, D-85748 Garching, Germany\\
$^2$School of Physics and Astronomy, University of Leeds, Leeds LS2 9JT, UK\\
$^3$Department of Physics and Astronomy, University of Victoria, PO Box 3055, STN CSC, Victoria, BC V8W 3P6, Canada\\
$^4$NRC-Herzberg Institute of Astrophysics, 5071 West Saanich Road, Victoria, BC V9E 2E7, Canada\\
$^5$Joint Astronomy Centre, 660 North A'ohoku Place, University Park, Hilo, HI 96720, USA\\
$^6$Department of Astrophysical and Planetary Sciences, University of Colorado, UCB 389 CASA, Boulder, CO 80389-0389, USA\\
$^7$Astronomy Department, University of California, 601 Campbell Hall MS 3411, Berkeley, CA 94720-3411, USA}

\maketitle

\begin{abstract}
Winds and supernovae from OB associations create large cavities in the interstellar medium referred to as superbubbles. The Orion molecular clouds are the nearest high-mass star-forming region and have created a highly elongated, 20$^{\circ}$ x 45$^{\circ}$, superbubble. We fit Kompaneets models to the Orion--Eridanus superbubble and find that a model where the Eridanus side of the superbubble is oriented away from the Sun provides a marginal fit. Because this model requires an unusually small scaleheight of 40 pc and has the superbubble inclined 35$^\circ$ from the normal to the Galactic plane, we propose that this model should be treated as a general framework for modelling the Orion--Eridanus superbubble, with a secondary physical mechanism not included in the Kompaneets model required to fully account for the orientation and elongation of the superbubble. 

\end{abstract}


\section{INTRODUCTION}
\label{intro}
		
	High-mass stars account for only a small fraction of the total mass of stars in the Galaxy, but dominate the energetics of the molecular clouds in which they form and significantly alter the conditions of the interstellar medium (ISM) in their vicinity (e.g. \citealt{Zinnecker07}). High-mass stars produce large fluxes of ionizing radiation that form large H{\sc ii} regions (e.g. \citealt{Shields90}). The strong winds of massive stars, coupled with the supernova blast waves created when these stars die, also create large, low-density cavities in the ISM that are filled with very hot gas (e.g. \citealt{Heiles76, McCray87, StaveleySmith97, Heyer98, Churchwell06, Churchwell07, Bagetakos11}). Extremely large, ionized bubbles that are formed from the collective energies of an OB association are referred to as superbubbles.
	 
	 The closest high-mass star-forming region to the Sun that is currently forming massive stars is the Orion star-forming region, containing the Orion molecular clouds, which is located at a distance of 400 pc from the Sun \citep{Hirota07,  Menten07,Sandstrom07}. The Orion star-forming region is surrounded by a highly elongated superbubble, with dimensions of 20$^\circ \times 45^\circ$ as seen in H$\alpha$ emission \citep{Bally08}, that is referred to as the Orion--Eridanus superbubble. Following the methodology of \citet{Basu99}, we attempt to fit the Orion--Eridanus superbubble with an analytic superbubble model to constrain the orientation and properties of the Orion--Eridanus superbubble.
	
	\citet{Kompaneets60} derived the structure of a bubble blown into an exponential atmosphere by a single explosion and this model has since been extended to bubbles driven by constant energy sources, such as stellar winds or series of supernova explosions (e.g. \citealt{Basu99}). Kompaneets models have been used to fit both observed \citep{Basu99, Pidopryhora07, Baumgartner13} and simulated superbubbles \citep{Stil09}, and have been shown to produce superbubble shapes consistent with the shapes produced by more complex models \citep{MacLow89}.  
	 	 
	  In Section \ref{Background}, we describe Kompaneets models in more detail and note the general features of the Orion--Eridanus superbubble. In Section \ref{fitting}, we fit Kompaneets models to the Orion--Eridanus superbubble and the quality of these fits is discussed in Section \ref{discussion}. The location of the ionization front within the superbubble is constrained in Section \ref{ion front}. Secondary processes not included in the Kompaneets model that might also influence the evolution of superbubble are discussed in Section \ref{secondary}, and we summarize our findings in Section \ref{conclusions}. We also include in Appendix \ref{properties} a summary of the physical properties of the superbubble derivable from our best-fitting Kompaneets model.
	  
\section{BACKGROUND}
\label{Background}

\subsection{Kompaneets Model}
\label{Kompaneets}

\citet{Kompaneets60} derived a semi-analytic solution for the propagation of a shock wave into a purely exponential atmosphere that is applicable to the evolution of a superbubble in the Galaxy's disc. This model has since been modified to allow for gas cooling \citep{Maciejewski99}; a time variable energy input rate \citep{Baumgartner13}; or a continuous injection of energy, as would be produced from stellar winds or a succession of supernovae \citep{Schiano85, Basu99}. A Kompaneets model has been used to successfully model numerous superbubbles, including the W4 superbubble \citep{Basu99, Baumgartner13}, the Ophiuchus superbubble \citep{Pidopryhora07}, and the local bubble \citep{Baumgartner13}. While more sophisticated numerical simulations of superbubbles exist \citep{Tomisaka86, MacLow88, BisnovatyiKogan89, MacLow89, TenorioTagle90, Tomisaka92, Tomisaka98, Stil09}, Kompaneets models have been shown to provide an excellent first-order solution for the shape of a bubble expanding in an exponential atmosphere (e.g. \citealt{MacLow89}). Kompaneets models are the current, de facto choice for analytic models of superbubbles. For this paper, we will use the continuous injection variant described in detail by \citet{Basu99} and will refer to such a model simply as a Kompaneets model. 

The three key assumptions of the Kompaneets model are (1) the pressure within the bubble is spatially uniform, (2) the bubble expands in the direction of the surface normal at all locations, and (3) the expansion speed of the bubble is given by Hugoniot jump conditions for a strong shock, where the exterior pressure is negligible compared to the interior pressure. The Kompaneets model predicts that the location of the wall of the superbubble is given by
\begin{equation}
r(z,y) = 2 H \arccos\left[\frac{1}{2} e^{z / 2H} \left(1 - \frac{y^2}{4H^2} + e^{-z/H}\right)\right],
\end{equation}
where $r$ is the perpendicular distance from the central axis of the superbubble to the bubble wall, $z$ is the distance along the central axis from the driving source, $H$ is the scaleheight of the exponential atmosphere, and $y$ is a parameter with dimensions of length which varies from 0 to $2H$ and essentially specifies the evolutionary stage of the bubble, with more evolved bubbles having a larger value of $y$. 

At early times in the evolution of a Kompaneets model, the superbubble remains relatively spherical and the $y$ parameter is roughly equal to the radius of the superbubble. After a bubble expands to a radial size approximately equal to the scaleheight of the exponential atmosphere, when $y \sim H$, the bubble begins to elongate significantly towards the lower density direction of the exponential atmosphere. Because of the constant pressure assumption of the Kompaneets model, at late times in the expansion of a Kompaneets model superbubble, the velocity of the top cap of the bubble becomes very large and the top cap reaches an infinite distance from the source in a finite time. For this paper, we will adopt the convention used by \citet{Basu99} and work with the dimensionless quantity $\tilde y = y/H$, since the shape of a superbubble is fully described by the value of $\tilde y$. 

To convert model size scales to physical units, one length scale must be specified. This length scale can be the scaleheight of the exponential atmosphere or the length of one feature of the bubble, specified by giving the distance and angular size of the feature. The pressure within the bubble, expansion velocity, time since the creation of the bubble, surface density of the bubble wall, interior temperature, driving luminosity, and initial density of the exponential atmosphere at the location of the driving source can all be given in physical units if two of these parameters are specified. Furthermore, the location of the ionization front can also be determined if, in addition to two of the above-mentioned parameters, the ionizing luminosity of the source and the temperature of the bubble wall are also provided. See also table 1 of \citet{Basu99} for the dimensional relationships between model parameters and appendix A of \citet{Basu99} for a more detailed description of Kompaneets models. 

\subsection{Orion--Eridanus Superbubble}
\label{superbubble}

	In this paper, all references to north or south refer to increasing or decreasing declination and references to east or west refer to increasing or decreasing right ascension, unless otherwise specified. References to above or below, however, will refer to directions along the major axis of the Orion--Eridanus superbubble either towards the lower density portion of the exponential atmosphere or the higher density portion, respectively. Thus, the top of the bubble will be the portion of the bubble farthest from the source and surrounded by lower density regions of the exponential atmosphere, and the bottom of the bubble will be the portion of the bubble closest to the source and surrounded by higher density portions of the exponential atmosphere. Similarly, we will use upper and lower to refer to directions along the major axis of the bubble. While `above' or `towards the top' will generally refer to larger absolute magnitudes of Galactic latitude, we emphasize that we intend these terms to refer to directions along the major axis of the superbubble and not a direction along a line of constant Galactic longitude, since the Orion--Eridanus superbubble's major axis appears to make an angle of approximately 30$^\circ$ to the normal to the Galactic plane in the plane of the sky.
		
	The Orion star-forming region is the nearest OB star-forming region to the Sun that is actively forming stars and is located 400 pc away in the constellation of Orion \citep{Hirota07, Menten07, Sandstrom07}. It is approximately 130 pc distant from the Galactic plane and contains both young subgroups of stars and two molecular clouds in which stars are actively forming (e.g. \citealt{Bally08}). 
		
	The Orion star-forming region is surrounded by prominent H$\alpha$ features that appear to trace out a 20$^{\circ}$ x 45$^{\circ}$ elongated superbubble structure, with enhanced H$\alpha$ flux within the confines of the bubble (e.g. \citealt{Pickering1890, Barnard1894, Elliott70, Elliott73, Isobe73, Reynolds74, Sivan74, Johnson78, Reynolds79, Boumis01, Haffner03}). The east side of the superbubble is dominated by a bright crescent of H$\alpha$ emission, which is referred to as Barnard's Loop \citep{Pickering1890, Barnard1894}. The west side of the superbubble is composed of a hook shaped H$\alpha$ feature, in the constellation of Eridanus, that can be broken into three arcs (e.g. \citealt{Meaburn65, Meaburn67,Johnson78}). Arc A is the eastern half of the hook, Arc B is the western half of the hook, and Arc C is the southern extension of the hook. For the remainder of this paper, when we refer to Arcs A, B, and C, we will be specifically referring to the H$\alpha$ emitting regions, and will refer to these three arcs collectively as the Eridanus filaments. The properties of the Eridanus filaments are discussed in more detail in \citet{Pon14a}.
		
\begin{figure}
   \centering
   \includegraphics[width=3in]{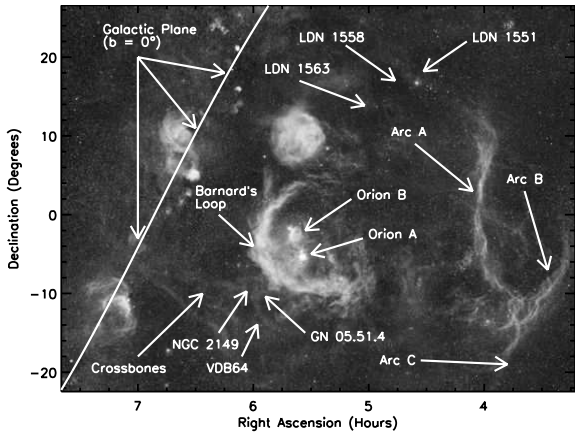}
   \caption{Orion--Eridanus superbubble as seen in H$\alpha$. Labels for the various major components of the bubble have been added to the image from \citet{DiCicco09}. Arcs A, B, and C are collectively referred to as the Eridanus filaments. The locations of additional small molecular clouds have been indicated, although many of these clouds are not obviously visible in the image, such that the arrows have been placed to point towards the known locations of the clouds. }
   \label{fig:diciccoalllabels}
\end{figure}

	The primary image used in this paper for determining the shape of the Orion--Eridanus superbubble is a wide field H$\alpha$ mosaic taken by astrophotographers \citep{DiCicco09}, which initially contained no specific astrometric information and was not calibrated to a known intensity scale. We identify 18 bright stars in the image and use their locations, as given in the {\it Hipparcos} data base \citep{Vanleeuwen07}, to astrometrically calibrate the image. This image is shown in Fig. \ref{fig:diciccoalllabels} and the most prominent H$\alpha$ features are labelled in this figure. 
		
	The Orion--Eridanus superbubble is also traced by a large H {\sc i} cavity \citep{Menon57, Heiles74Habing, Heiles76, Green91, Burrows93, Brown95, Kalberla05}, although the H {\sc i} cavity extends beyond the apparent H$\alpha$ limits of the superbubble, as might be expected if the ionizing photons from the Orion star-forming region fully penetrate the walls of the superbubble. This larger H {\sc i} extent is also seen in the Ophiuchus superbubble, which has been successfully fit by a Kompaneets model \citep{Pidopryhora07}.
		
\begin{figure}
   \centering
 \includegraphics[width=3in]{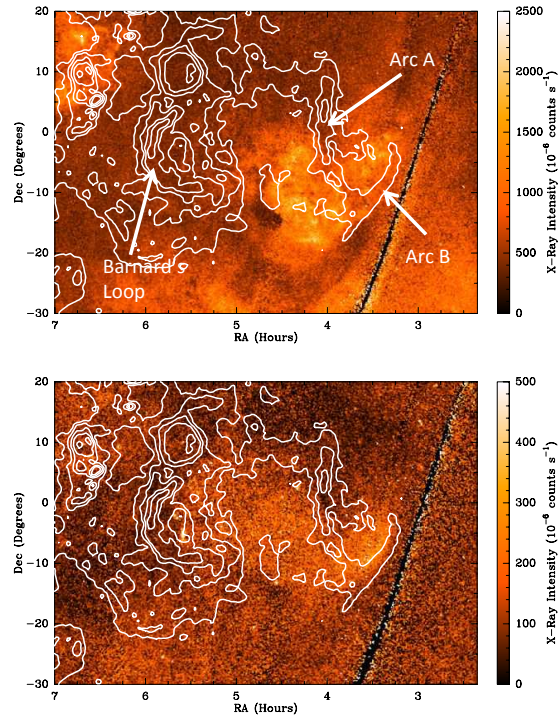}
   \caption{Diffuse X-ray emission observed by the {\it ROSAT} satellite \citep{Snowden97}. The top panel shows the 0.25 keV, R12 band and the bottom panel shows the 0.75 keV, R45 band. The contours are integrated intensities of H$\alpha$ from the WHAM survey \citep{Haffner03} and are logarithmically spaced with each contour corresponding to a factor of 2 increase in intensity. The lowest contour corresponds to an integrated intensity of 10 R. The locations of Arc A, Arc B, and Barnard's Loop are labelled.}
   \label{fig:rosat}
\end{figure}

	Rocket- and satellite-based X-ray surveys show that there is a diffuse enhancement of soft, 0.25 and 0.75 keV, X-rays towards the Orion--Eridanus superbubble \citep{Davidsen72, Williamson74, Naranan76, Long77, Fried80, Nousek82, Singh82, McCammon83, Marshall84, Garmire92, Burrows93, Guo95, Snowden95Burrows, Snowden97}. This X-ray emission comes from plasma with a temperature of approximately $2 \times 10^6$ K \citep{Burrows93}, as would be expected for the hot interior of a superbubble. Fig. \ref{fig:rosat} shows the 0.25 and 0.75 keV bands of the {\it R\"{o}ntgensatellit} ({\it ROSAT}) mission, which are the R12 and R45 bands, respectively, of \citet{Snowden97}. 

The enhanced 0.75 keV X-ray emission extends from the Orion star-forming region to Arc B and does not significantly extend north or south of the H$\alpha$ declination limits of the superbubble or to the east of Arc B, suggesting that the H$\alpha$ does indeed trace the superbubble walls. There is, however, some 0.25 keV emission extending to the south of the superbubble's southernmost H$\alpha$ boundary. The 0.25 keV emission partially fills a H {\sc i} void extending south to a declination of -20$^{\circ}$, between right ascensions of 4$^h$ and 4$^h$30$^m$, with the X-ray emission preferentially filling the north-west side of the H {\sc i} extension. This southern extent of the 0.25 keV emission has been suggested as being due to a separate bubble blown by a single star \citep{Burrows93}, a small-scale blowout of the superbubble \citep{Heiles99}, or the diffuse hot halo background \citep{Snowden95Burrows}. The leakage of hot gas from the interior of a bubble appears to be common for evolved H {\sc ii} regions \citep{Lopez13}. 

A more detailed summary of prior observations of the superbubble is presented in \citet{Pon13Thesis}.

\subsection{Validity of Applying a Kompaneets Model}
\label{validity}

The Kompaneets model is one of the most widely used analytic models for analysing superbubbles (e.g. \citealt{Basu99, Pidopryhora07,Baumgartner13}). While there are some properties of the Orion--Eridanus superbubble that are not immediately obviously consistent with a Kompaneets model, these properties are all seen in other superbubbles that have been successfully fit with a Kompaneets model, such as the Ophiuchus superbubble, the local bubble, and the W4 superbubble.

The major axis of the Orion--Eridanus superbubble is inclined $30^\circ$ with respect to the normal to the Galactic plane in the plane of the sky. The Ophiuchus superbubble and the local bubble are also inclined from the normal to the Galactic plane, although by only $15^\circ$ \citep{Pidopryhora07} and $20^\circ$ \citep{Baumgartner13}, respectively. While many models of the Orion--Eridanus superbubble have the bubble elongated close to parallel to the Galactic plane (e.g. \citealt{Reynolds79, Burrows93,Burrows96, Lee09}), the angle between the normal to the Galactic plane and the major axis of the superbubble is dependent upon the inclination of the superbubble with respect to the plane of the sky. We investigate various orientations of the bubble with respect to the plane of the sky, such that we can obtain Kompaneets model fits that are only inclined by 35$^{\circ}$ with respect to the normal to the Galactic plane.

The large elongation of the Orion--Eridanus superbubble requires a $\tilde y$ value close to 2, since significant elongation only occurs once $\tilde y > 1$. Such large $\tilde y$ values are, however, also required for the W4 superbubble, Ophiuchus superbubble, and the local bubble, with these three bubbles requiring $\tilde y$ values of 1.96 - 1.98 \citep{Basu99, Baumgartner13}, 1.96 - 1.998 \citep{Pidopryhora07}, and 1.8 \citep{Baumgartner13}, respectively.

Since the nearest superbubble to the Orion--Eridanus superbubble, the local bubble, has been successfully fit with a Kompaneets model, it suggests that the ISM into which the Orion--Eridanus superbubble is expanding might be well described as an exponential atmosphere. The gas layer of the Galactic disc that is most relevant for the Orion--Eridanus superbubble is the cold gas layer, which has a scaleheight of 100-150 pc \citep{Kalberla09}. The warm neutral H {\sc i}, warm ionized gas, and hot highly ionized gas layers of the Galactic disc have scaleheights of 500 pc, 1.5 kpc, and 4.4 kpc, respectively, which are all larger than the extent of the Orion--Eridanus superbubble  \citep{Lockman84, Reynolds89, Savage97}. 

The presence of a giant molecular cloud, out of which the stars powering a superbubble form, adds material above that expected from an exponential atmosphere and can hinder the expansion of a superbubble, particularly towards the Galactic plane. While the lower half of the W4 superbubble has a smaller size than predicted by Kompaneets models due to the extra mass associated with the W4 loop, a Kompaneets model fits the upper part of the superbubble very well \citep{Baumgartner13}. Models of superbubbles expanding in Gaussian atmospheres show that superbubbles rapidly grow in radius to match the effective scaleheight of each layer of the atmosphere the bubble enters \citep{MacLow89}, such that any extra material present in the Orion star-forming region should have little effect on the shape of the Orion--Eridanus superbubble's Eridanus half. 			
			
\section{MODEL FITTING}
\label{fitting}

\subsection{Selection of Components}
\label{selection}

As described by a Kompaneets model, a superbubble's wall is particularly efficient at absorbing ionizing photons from the source due to the higher density of the wall compared to the interior of the bubble or the gas exterior to the bubble. Thus, the Kompaneets model predicts that the edge of a superbubble should be well defined by enhanced H$\alpha$ emission, as H$\alpha$ emission is produced from the recombination of previously dissociated protons and electrons. For superbubbles expanding in exponential atmospheres, most of the ionizing photons that escape through the edge of the superbubble wall are capable of travelling large distances from the superbubble before being absorbed, such that H {\sc i} emission is expected to only trace the edge of the superbubble wall where the ionization front is trapped within the wall \citep{Basu99}. Thus, the Kompaneets model predicts that H$\alpha$ should be a better tracer of the morphology of a superbubble than H {\sc i}, and we choose to use the H$\alpha$ extent of the Orion--Eridanus superbubble to fit Kompaneets models to. 

Because of the relatively small distance between the Orion--Eridanus superbubble and the Sun, there are numerous small H$\alpha$ features visible in the vicinity of the Orion--Eridanus superbubble and it is non-trivial to select which features are associated with the bubble wall and which features are just small density fluctuations in the ISM illuminated by ionizing photons leaking out of the superbubble. We use the 0.75 and 0.25 keV X-ray emission to inform our decision of which H$\alpha$ features correspond to the edge of the superbubble, as well as trying to preferentially choose features closer to the centre of the superbubble, as the Kompaneets model predicts no significant internal structure to a superbubble. Therefore, we select the H$\alpha$ filament running along a declination of approximately -12$^\circ$ as the southern extent of the superbubble and the filament running from the north end of Barnard's Loop to the top of Arc A, at (4$^h$0$^m$, 10$^\circ$), as the northern edge of the superbubble. We therefore select Arcs A and B as being part of the bubble wall, but place Arc C outside of the bubble wall. It is still quite plausible that Arc C is related to the superbubble, but has been formed by a process not included in the Kompaneets model, such as a spatially localized blow out of hot gas from the bubble wall (e.g. \citealt{Heiles99}). We have chosen the bubble wall to contain the regions where both 0.75 and 0.25 keV X-ray emission are observed, but not the southern extent of the 0.25 keV emission where no 0.75 keV emission is detected. This is because the nature of this southern extension of the 0.25 keV emission is not well agreed upon \citep{Burrows93, Snowden95Burrows, Heiles99}.

\subsection{Model T: towards the Sun}
\label{near}
	
	Studies of interstellar absorption features in stellar spectra towards the Eridanus half of the bubble reveal a wall of gas moving towards the Sun at a distance of approximately 180 pc, which has often been interpreted as being the near wall of the superbubble \citep{Frisch90,Guo95, Burrows96,Welsh05}. With this interpretation, we first attempt to fit a Kompaneets model in which the superbubble is oriented with the Eridanus end pointed towards the Sun and where the near side of the bubble is approximately 180 pc distant. The free parameters for this fit are the position of the easternmost point of the bubble, the orientation of the major axis of the bubble in the plane of the sky, the elongation of the bubble (as controlled by the $\tilde y$ parameter), the inclination of the bubble with respect to the plane of the sky, and the maximum width of the bubble. For all fits, the Orion end of the superbubble is assumed to be 400 pc away, corresponding to the estimated distance of the Orion star-forming region \citep{Menten07}. The scaleheight of the ISM (the cold H {\sc i}) exterior to the bubble and parallel to the major axis of the bubble is also obtained from the model fit. Because of the complex structure of the region, all fitting is done by eye. We estimate the relative error of this fitting below. 
	
	The location of the driving source is not used as an input parameter, but rather, is derived from the model. We do not use the driving source as an input parameter because it is not clear which of the four separate subgroups of stars that have formed in the last 10 Myr towards Orion \citep{Brown94} have significantly contributed to the formation of the Orion--Eridanus superbubble. Furthermore, the location of the driving source has likely shifted slightly over time, due to the proper motions of the various subgroups in Orion. Fits of Kompaneets models to the W4 superbubble also show that its driving source, OCl 352, is offset from the required source location for the best-fitting Kompaneets model.
		
\begin{figure}
   \centering
 \subfigure{
      \includegraphics[width=3in]{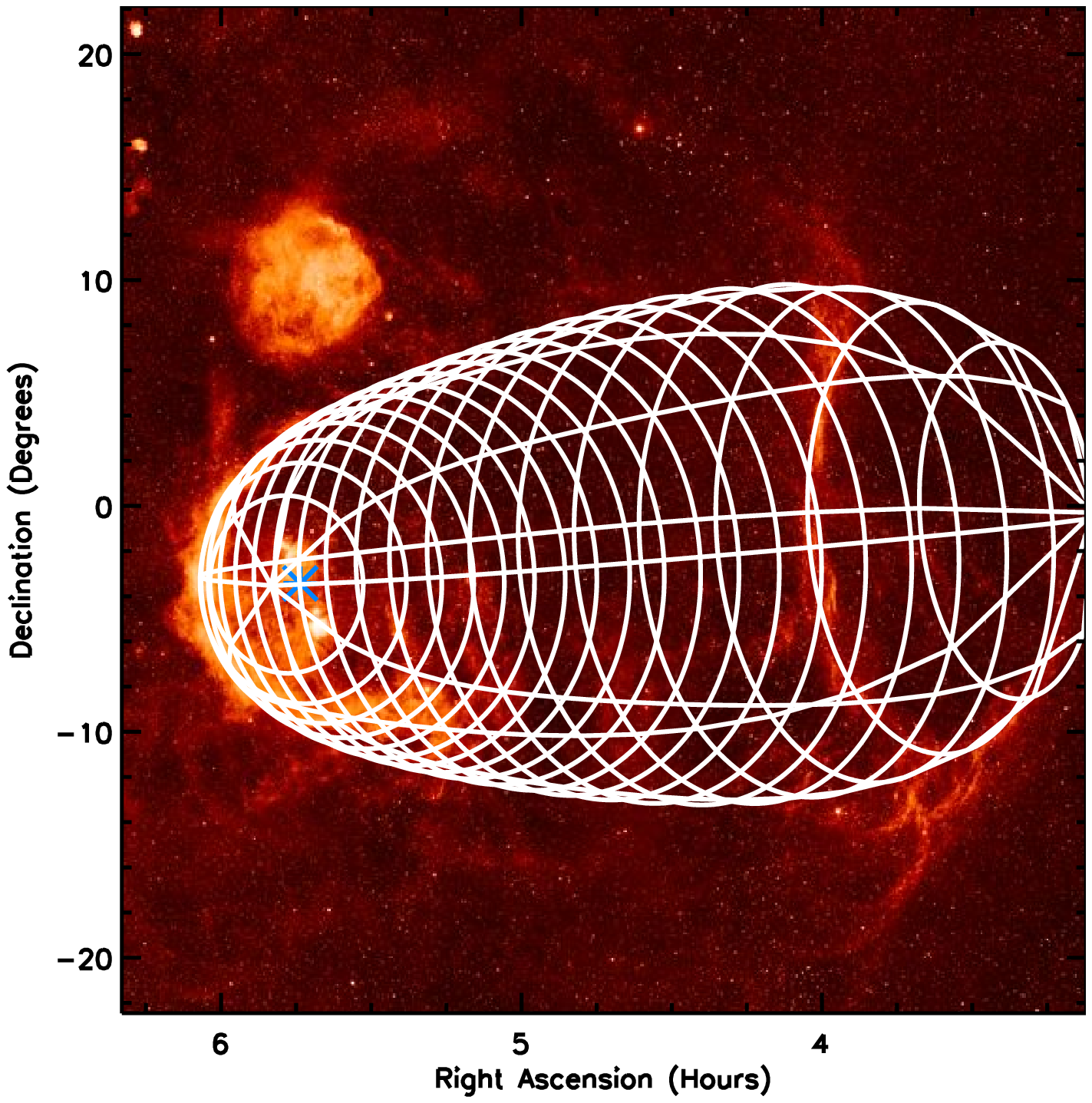}}
 \subfigure{
      \includegraphics[width=3in]{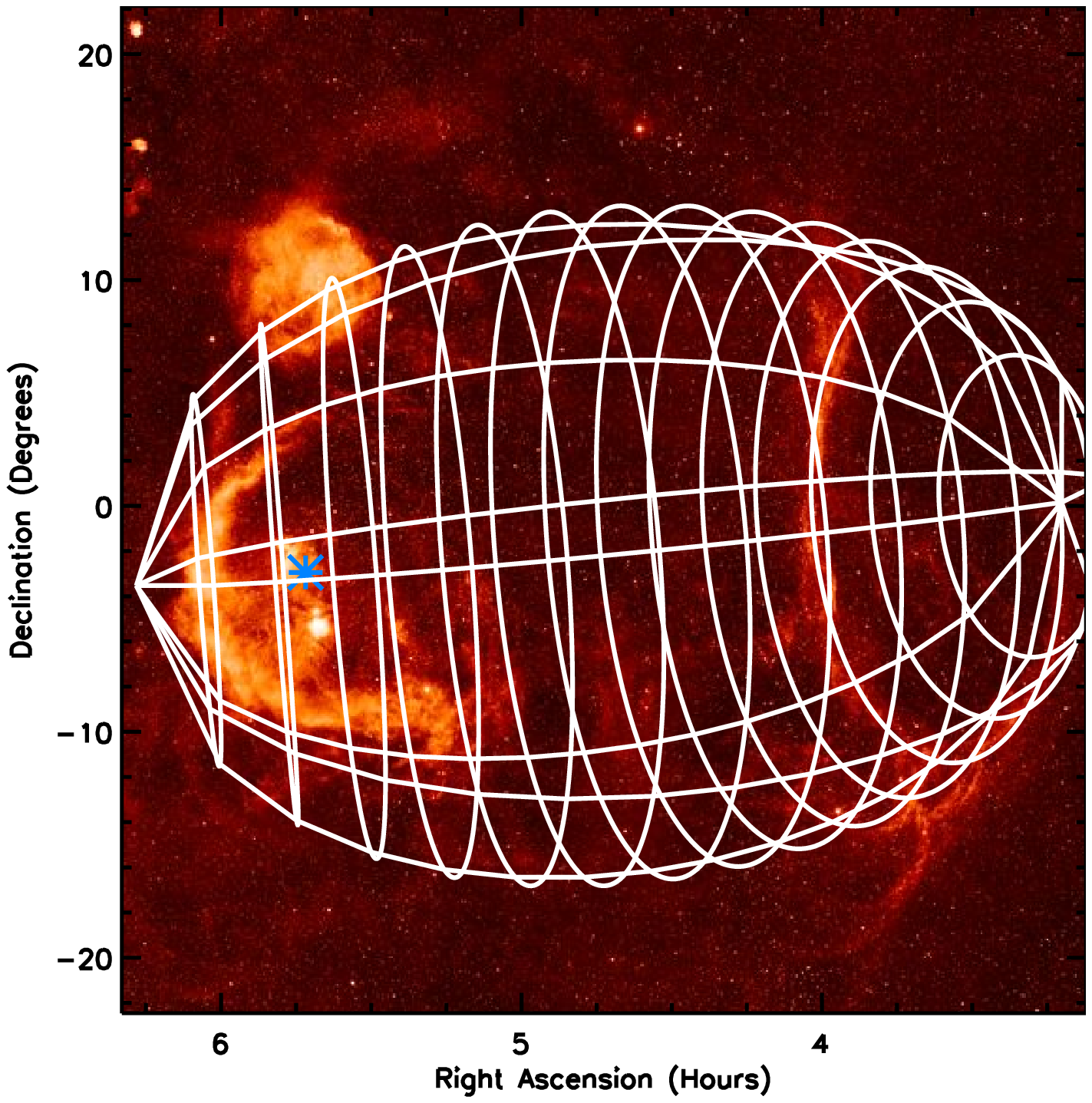}}
   \caption{Best-fitting Kompaneets models of the Orion--Eridanus superbubble, white lines, overplotted on the H$\alpha$ image of \citet{DiCicco09}. The top panel shows the best-fitting model for a bubble oriented towards the Sun, model T, while the bottom panel shows the best-fitting model for a bubble oriented away from the Sun, model A. The asterisks denote the locations of the driving sources in the Kompaneets models and bubble parameters are given in Table \ref{table:komp properties}.}
    \label{fig:nearandfarbubble}
\end{figure}

\begin{figure}
   \centering
 \subfigure{
      \includegraphics[width=3in]{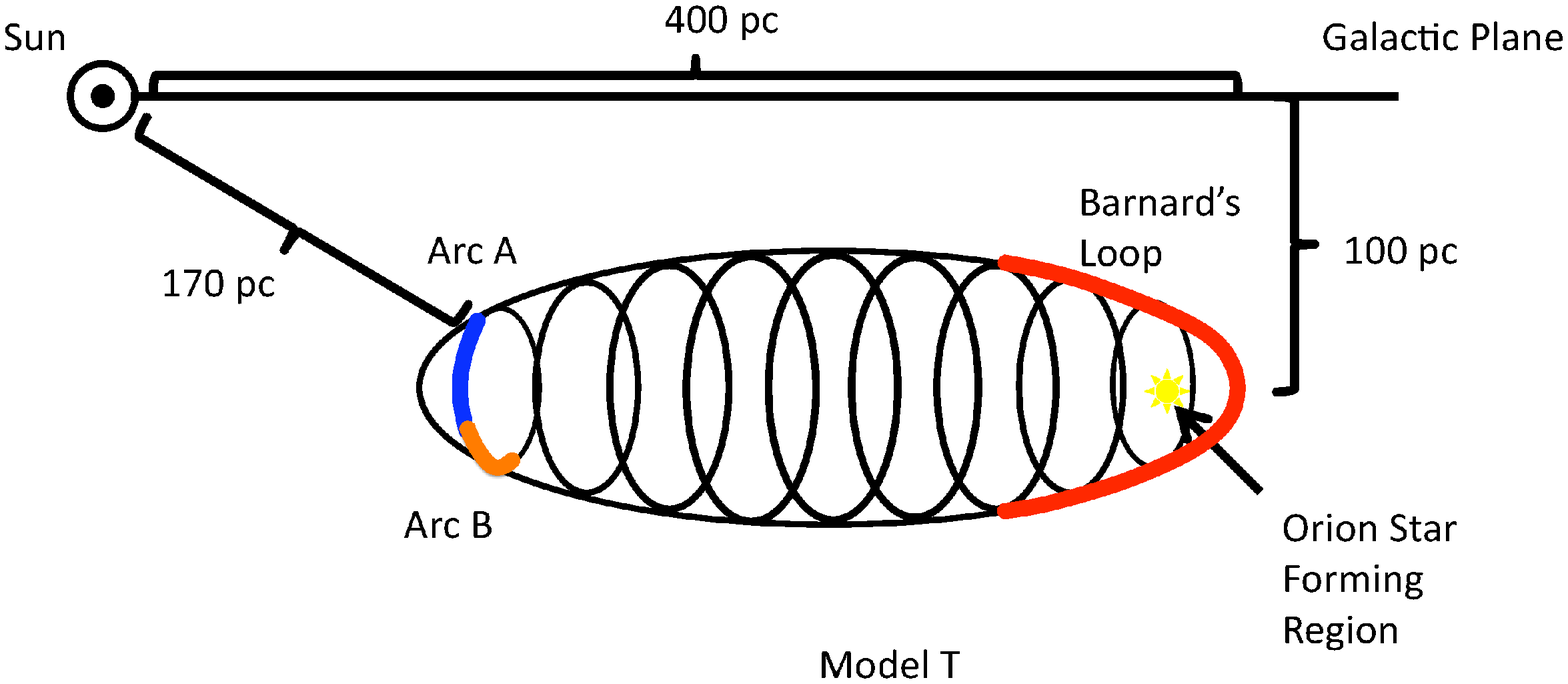}}
 \subfigure{
      \includegraphics[width=3in]{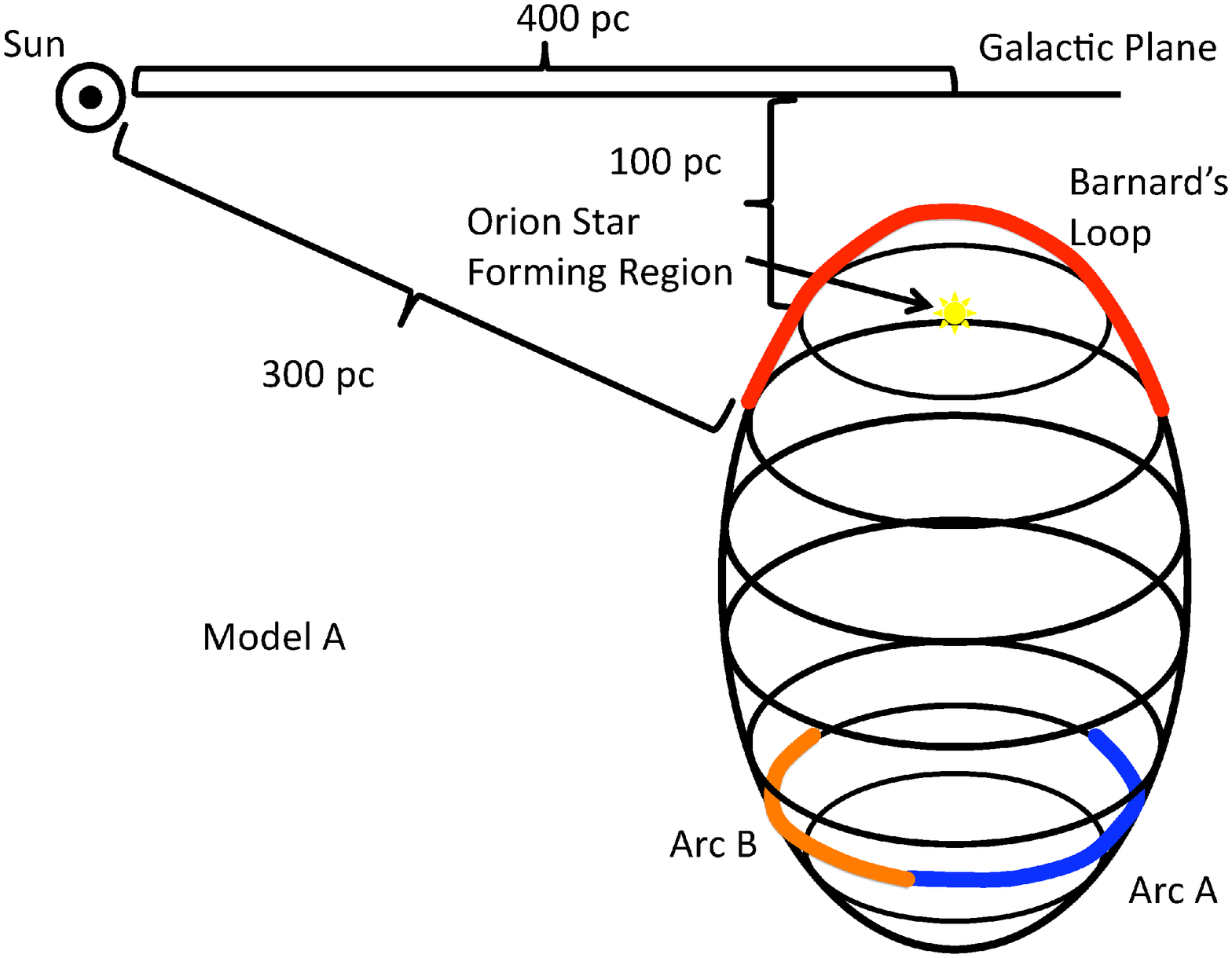}}
   \caption{Schematic diagram of models T and A. Model T is shown in the top panel and model A is shown in the bottom panel. In each of these schematic diagrams, the superbubble major axis is inclined roughly 30$^\circ$ into the page, in order to match the offset between the superbubble axis and the normal to the Galactic plane apparent in Fig. \ref{fig:diciccoalllabels}. Thus, the major axis of the bubble in model A is still 35$^\circ$ inclined from the normal to the Galactic plane.}
    \label{fig:schematic}
\end{figure}

	The best-fitting model for this bubble orientation, in which the closest point on the near side of the bubble is 171 pc from the Sun, is shown in the top panel of Fig. \ref{fig:nearandfarbubble} and will hereafter be referred to as model T (for `towards'). On the top panel of Fig. \ref{fig:nearandfarbubble}, the position of the driving source, as predicted by the Kompaneets model, is denoted with an asterisk and is roughly coincident with the position of the Orion B molecular cloud. A schematic diagram of this model is shown in the top panel of Fig. \ref{fig:schematic} and the basic parameters of this model are given in Table \ref{table:komp properties}. 
	
	Reasonably good fits can also be obtained, although not shown, where Arc B corresponds to the end cap of the bubble and Arc A is a feature on the near side of the bubble. Because the fits are tightly constrained by the size of Barnard's Loop and the requirement that the near side be approximately 180 pc distant, no adequate fits are found with a value of $\tilde y$ below 1.999 or with a scaleheight greater than 20 pc.
		
\begin{table}
\caption{Kompaneets model properties}
\begin{center}
\begin{tabular}{cccccc}
\hline
Model & $\tilde y$ & $H$ (pc) & $d_{min}$ (pc) & $d_{max}$ (pc) & $\theta$ ($^\circ$)\\
(1) & (2) & (3) & (4) & (5) & (6)\\
\hline
T & 1.9995 & 15 & 170 & 400 & 85\\
A & 1.945 & 40 & 305 & 540 & 35\\
\hline
\end{tabular}
\end{center}
{\it Note.} Column (1) gives the model name. Columns (2) and (3) give the value of the $\tilde y$ parameter and the scaleheight of the exponential atmosphere, respectively. Columns (4) and (5), respectively, give the distance to the closest and farthest point on the superbubble wall. The maximum distance in model T occurs at the Orion end while the far side of the bubble near the location of Arc A is only 275 pc distant. Column (6) gives the angle that the major axis of the superbubble makes with the normal to the Galactic plane.
\label{table:komp properties}
\end{table}
			
\subsection{Model fit A: away from the Sun}
\label{far}
	
	There is some recent evidence that Arc A and the back side of the Orion--Eridanus superbubble are located at a distance greater than 500 pc \citep{Boumis01, Welsh05, Ryu06}. For a more thorough discussion on the possible distance to Arc A, see \citet{Pon14a}. The absorption studies that place the near side of the superbubble at 180 pc also use lines coming from neutral, rather than ionized gas and thus, these studies do not necessarily trace the ionized wall of the bubble and may give a near distance that is too small. Thus, we attempt to find a second model fit where the bubble is oriented away from the Sun, such that Arc A resides on the far side of the bubble and we drop the constraint that the near side of the bubble must be roughly 180 pc distant from the Sun. For this model orientation, we find that Barnard's Loop cannot be accurately fit and thus the position of the driving source is used as a fitting parameter, rather than the location of the easternmost part of the bubble, as was done for model T. As before, the other fitting parameters used are the orientation of the major axis of the bubble in the plane of the sky, the elongation of the bubble, the inclination of the bubble with respect to the plane of the sky, and the maximum width of the bubble. The source location is constrained to lie near the Orion B molecular cloud and the far eastern edge of the bubble is assumed to be at a distance of 400 pc. Once again, all fitting is done by eye.

	The best-fitting model for this orientation is shown in the bottom panel of Fig. \ref{fig:nearandfarbubble} and will hereafter be referred to as model A (for `away'). A schematic diagram of this model is given in the bottom panel of Fig. \ref{fig:schematic} and the basic parameters of this model are given in Table \ref{table:komp properties}. 
	
	 A slight reduction in the $\tilde y$ parameter also produces a reasonable fit where Arc B is the end cap of the bubble. Because there is no tight distance constraint on the Eridanus end and no constraints on the shape of the Orion end, a much wider array of models produce reasonable fits for a bubble oriented away from the Sun. All reasonable bubble fits, however, have $\tilde y > 1.7$ and scaleheights less than 60 pc. In model A, Arc A is located at a distance of 500 pc and the closest point on the near side of the bubble is just over 300 pc distant from the Sun. The maximum diameter of the bubble is 220 pc in model A, as compared to 100 pc in model T. For a discussion of other physical parameters derivable from this model, please see Appendix \ref{properties}.
	 
\section{DISCUSSION}
\label{discussion}

	Model T does a reasonable job of reproducing the morphology of the Orion--Eridanus superbubble. In this model, both Arcs A and B lie along the same plane perpendicular to the major axis of the superbubble, such that the filaments trace the bubble wall at a particular, constant distance from the driving source. Given the orientation of the bubble towards the Sun, Arc A traces the near side of the bubble while Arc B traces the far side of the bubble. The sense of curvature of Arc A comes out naturally from this model fit. 
	
	Unfortunately, this model has two significant flaws. First, to match the observed elongation of the superbubble, model T requires a scaleheight of only 15 pc for the Galactic disc's atmosphere. The scaleheight of the cold gas layer of the Galaxy is generally estimated to be closer to 100-150 pc \citep{Kalberla09}, a factor of 10 larger than the scaleheight in model T. This 15 pc scaleheight of model T is even smaller than the 25 and 27 pc scaleheights derived from fits to the W4 superbubble \citep{Basu99, Baumgartner13}. 
	
	Second, to match the near side distance, model T is 85$^\circ$ inclined with respect to the normal to the Galactic plane. This is completely at odds with the expected direction of the gradient in the Galactic ISM density profile and thus we reject Model T as being an adequate fit of the Orion--Eridanus superbubble.
	
	We caution, however, that while we believe that Model T is not an adequate fit of the Orion--Eridanus superbubble, we do not rule out the possibility that the near side of the superbubble is 180 pc distant and the superbubble is elongated parallel to the Galactic plane. We merely conclude that if the superbubble is elongated parallel to the plane, then the evolution of the bubble is not consistent with it being controlled by pressure-driven expansion into the typically assumed exponential atmosphere of the Galactic disc. 
	
	An orientation parallel to the Galactic plane is not completely unheard of for a superbubble. Catalogues of H{\sc i} shells and supershells show that these shells tend to be preferentially elongated parallel to the Galactic plane \citep{Heiles79, Ehlerova05, Ehlerova13, Suad14}. For instance, of the 190 H{\sc i} supershells studied by \citet{Suad14}, for which the mean weighted eccentricity is 0.8, approximately 70\% of the supershells are oriented such that their major axis is closer to being parallel to the Galactic plane than being perpendicular.

	Model A does not fit Barnard's Loop, nor any of the weak emission to the north of the bubble, but does fit some weak features seen to the south of the bubble and does a reasonable job fitting the Eridanus filaments. As in model T, Arc B and Arc A both lie in the same plane perpendicular to the major axis of the bubble at a fixed distance from the driving source. 

	The failure of model A to reproduce the size of Barnard's Loop is not completely unexpected. The Kompaneets model assumes that the atmosphere into which a bubble is expanding is a pure exponential atmosphere and does not take into account the presence of any density enhancements, such as that provided by a giant molecular cloud. The greater density and pressure of a molecular cloud would hinder the expansion of a superbubble in the direction of that cloud, as has been suggested for the W4 superbubble \citep{Baumgartner13}. Thus, if the bulk of the molecular cloud complex that gave rise to the Orion star-forming region were originally located to the east of the driving source of the superbubble, it would be expected that the observed size of the superbubble in this direction would be smaller than predicted by a Kompaneets model. The failure of model A to match the size of Barnard's Loop might simply be due to the expansion of the superbubble being slowed by the excess material present in and around the Orion star-forming region. 
		
	The vertical scaleheight of the exponential atmosphere in model A is 40 pc. This is still significantly smaller than the typical scaleheight of the Galactic disc, but is closer to the accepted value than either the scaleheight required for model T or the scaleheight derived for the W4 superbubble \citep{Basu99, Baumgartner13}. The small scaleheights of models A and T are motivated by the observed elongation of the Orion--Eridanus superbubble, because, in a Kompaneets model, a bubble remains roughly spherical until the bubble has expanded to a radial size of a few scaleheights, as described in Section \ref{Kompaneets}. 
	
	It is plausible, albeit unlikely, that the gas in the vicinity of the Orion--Eridanus superbubble has been sufficiently vertically compressed in the past to explain this small scaleheight. Some compression is expected for overdensities in the disc due to the extra gravitational force of the overdensities. The Parker instability \citep{Parker66} can also further decrease the scaleheight of the ISM. 
	
	Model A has the major axis of the superbubble making an angle of 35$^\circ$ with respect to the normal to the Galactic plane. While not perfectly perpendicular to the Galactic plane, the major axis of model A is much closer to alignment with the normal to the Galactic plane than model T. Given that the projection of the major axis of the superbubble on to the plane of the sky makes a 30$^\circ$ angle with the normal to the Galactic plane, model A is oriented such that the major axis of the bubble is almost as close to perpendicular to the Galactic plane as possible. 
	
	It is plausible that a 35$^\circ$ difference in orientation could be due to a secondary process shaping the superbubble in conjunction with the elongation induced by the exponential atmosphere of the Galactic disc. We thus adopt model A as our preferred model for understanding the Orion--Eridanus superbubble. We believe that the expansion of the superbubble in the Galaxy's exponential atmosphere, coupled with some secondary process that slightly collimates the bubble, could easily produce a structure consistent with what is observed, with the superbubble oriented away from the Sun. 

\subsection{Detailed Summary of Quality of Models A and T}
\label{quality}

In the following subsection, we quickly summarize the detailed advantages (+) and disadvantages (--) of models T and A. The advantages and disadvantages of model T are the following.
\begin{itemize}
\item[+] The model accurately traces both Barnard's Loop and the Eridanus filaments and the driving source in the model is consistent with the location of the Orion B molecular cloud.
\item[+] The near side of the model is approximately 180 pc away, consistent with the detection of a wall of gas with negative velocities at this distance (e.g. \citealt{Welsh05}).
\item[+] Arc A is on the near side of the bubble and thus could produce the absorption features seen in diffuse X-ray emission (e.g. \citealt{Snowden95Burrows}), while Arc B is on the far side of the bubble, which is expected since the typical velocity associated with Arc B is similar to that of the larger of two velocity components detected towards the Eridanus half of the bubble \citep{Reynolds79}.
\item[-] The scaleheight required to fit the elongation of the bubble, 15 pc, is much smaller than the generally accepted scaleheight for the cold H {\sc i} in the Galactic disc, 100-150 pc \citep{Kalberla09}. 
\item[-] The bubble is elongated almost parallel to the Galactic plane, which is perpendicular to the expected density gradient in the disc.
\item[-] The model places Arc A on the near side of the bubble and Arc B on the backside, such that Arc A should have a more negative velocity than Arc B, whereas observations reveal that Arc A has a more positive velocity \citep{Haffner03}.
\item[-] The far side of the bubble in the model is less than 300 pc distant, whereas absorption studies have not detected the far side within 500 pc (e.g. \citealt{Welsh05}). 
\item[-] Arc A is 200 pc distant, which is inconsistent with the upper limits for the proper motion of Arc A derived by \citet{Boumis01}.
\item[-] The bubble orientation is not well aligned with the local magnetic field. 
\end{itemize}

The advantages and disadvantages of model A are the following.
\begin{itemize}
\item[+] The model accurately traces the Eridanus filaments.
\item[+] The far side of the model, as well as Arc A, is over 500 pc distant, consistent with non-detections in absorption and proper motion studies (e.g. \citealt{Boumis01, Welsh05}).
\item[+] Arc B is located closer to the Sun than Arc A, as suggested from radial velocities \citep{Haffner03}.
\item[+] The major axis of the bubble is more closely aligned with the normal to the Galactic plane, with the major axis being $35^\circ$ offset from the normal. 
\item[+] The projection of the bubble on to the Galactic plane is in the direction of the local magnetic field.
\item[-] The scaleheight of the model, 40 pc, is smaller than expected.
\item[-] The model does not reproduce the shape of Barnard's Loop.
\item[-] The near side of the bubble is over 300 pc distant and thus this model cannot explain the absorption features detected in the spectra of stars with distances greater than 180 pc (e.g. \citealt{Welsh05}). 
\item[-] Arc B is on the near side of the bubble, opposite of what is suggested from radial velocities \citep{Haffner03}, and Arc A is on the far side of the bubble, such that it cannot be the cause of absorption features seen in diffuse X-ray surveys (e.g. \citealt{Snowden95Burrows}). 
\end{itemize}
				
\section{IONIZATION FRONT FITTING}
\label{ion front}

The location where an ionization front breaks out of a superbubble can be observationally determined by locating where H {\sc i} emission disappears from the bubble wall, as done in \citet{Basu99}. 

Because H$\alpha$ emission is powered by the absorption of ionizing photons, H$\alpha$ intensity should be proportional to the energy of ionizing radiation absorbed. Where the ionization front occurs within the bubble wall, the H$\alpha$ surface brightness should vary roughly with the flux of incident ionizing photons, as seen from the ionizing source, since all of the ionizing photons are absorbed within the wall. As such, the H$\alpha$ flux in these regions should vary inversely with the square of the distance from the source. For regions where the ionization front lies outside of the superbubble, the H$\alpha$ intensity should depend upon both the ionizing flux reaching the wall and the fraction of the ionizing flux that the wall captures. Thus, the H$\alpha$ brightness of the wall should show a discontinuity where the ionization front breaks out, as the H$\alpha$ intensity above the point where the ionization front breaks out should be lower than predicted from the inverse square distance dimming seen below the ionization front breaking out. 

\begin{figure}
   \centering
   \includegraphics[width=3 in]{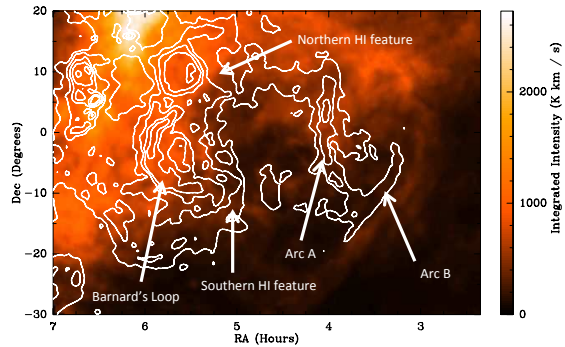}
   \caption{H {\sc i} integrated intensities in the range $-10 \le V_{\mbox{LSR}} \le 20$ km s$^{-1}$ from the LAB Galactic H {\sc i} Survey are shown in the colour scale. The contours are integrated intensities of H$\alpha$ from the WHAM survey \citep{Haffner03} and are logarithmically spaced with each contour corresponding to a factor of 2 increase in intensity. The lowest contour corresponds to an integrated intensity of 10 R. The H$\alpha$ integrated intensities are velocity integrated between -80 and 80 km s$^{-1}$. The locations of Arc A, Arc B, and Barnard's Loop are labelled. The northern and southern H {\sc i} features, which we interpret as indicating the locations where the ionization front breaks out of the Orion--Eridanus superbubble, are also labelled.}
   \label{fig:labintn10p20}
\end{figure}

	There is a noticeable linear H {\sc i} feature extending from the northern tip of Barnard's Loop. The  H {\sc i} cavity associated with the Orion--Eridanus superbubble also begins to extend to the north of the H$\alpha$ boundary of the bubble around the north-western edge of Barnard's Loop. Similarly, the H{\sc i} cavity extends to the south of the H$\alpha$ limits of the superbubble just past the south-western tip of Barnard's Loop. As such, these H {\sc i} data seem to indicate that the ionizing photons from the Orion star-forming region are breaking out of the superbubble at the western edges of Barnard's Loop. This location would also be consistent with the significant drop in H$\alpha$ flux at the western edges of Barnard's Loop and the diffuse H$\alpha$ emission coming from outside of the H$\alpha$ features we have identified as being the bubble wall. Fig. \ref{fig:labintn10p20} shows the integrated intensity of 21-cm emission, for $-10 \le V_{\mbox{LSR}} \le 20$ km s$^{-1}$, from the Leiden/Argentine/Bonn (LAB) Galactic H {\sc i} Survey \citep{Hartmann97, Arnal00, Bajaja05, Kalberla05}, with the Wisconsin H-Alpha Mapper (WHAM; \citealt{Haffner03}) H$\alpha$ integrated intensities overplotted as contours and arrows pointing to the northern and southern H {\sc i} features extending from Barnard's Loop.
	
	LDN 1551 lies to the north of the Orion--Eridanus superbubble, given the bubble boundary in both models T and A, and slightly to the west of Barnard's Loop. \citet{MoriartySchieven06} examine this small molecular cloud and find evidence that it has been partially photoevaporated by the Orion star-forming region. \citet{Lee09} suggest that LDN 1551, as well as LDN 1558 and LDN 1563 for similar reasons, must be on the edge of the Orion--Eridanus superbubble for the clouds to be influenced by the ionizing radiation from the Orion star-forming region. We instead suggest that these clouds lie outside the superbubble, based upon the confinement of the 0.75 keV X-ray emission, but are still irradiated by ionizing radiation from the Orion star-forming region that is penetrating the bubble wall past the edge of Barnard's Loop. That is, we suggest that the apparent photoevaporation of these clouds is further evidence that the ionizing photons from the Orion star-forming region breakout of the superbubble to the west of Barnard's Loop. These clouds may also have been significantly shaped during prior intervals of time when the ionizing flux from the Orion star-forming region was greater than it is now. Such variations in flux over time would be expected as the stellar population of the Orion star-forming region changes. 
	
	Barnard's Loop may also be slightly porous to ionizing photons as NGC 2149, GN 05.51.4, VDB64, and the Crossbones all show evidence of being affected by the superbubble, despite Barnard's Loop being between these small molecular clouds and the Orion star-forming region on the plane of the sky \citep{Lee09}. We note that soft, non-ionizing ultraviolet (UV) photons that penetrate the superbubble wall can also help shape clouds outside of the superbubble. Alternatively, projection effects may only make it appear that there are no direct lines of sight between the Orion star-forming region and these small clouds. Fig. \ref{fig:diciccoalllabels} shows the location of all of the above-mentioned clouds.

	As shown by \citet{Basu99}, it is possible to theoretically calculate whether, and where, an ionization front will break-out of a particular Kompaneets model, given the initial density of the ambient gas at the location of the source, the ionizing photon flux of the source, the temperature of the material in the bubble wall, and the gas pressure of the bubble wall. An ionization front will remain within a bubble wall if the depth required to absorb all of the ionizing photons is larger than the line of sight depth of the bubble wall, such that
\begin{equation}
\left(\frac{\Phi_* \, \cos(\phi) \, k T}{2 \pi \, N_s \, P_s \, \alpha_b \, s^2}\right) > 1,
\label{eqn:ionfrontcriteria}
\end{equation}
where $\Phi_*$ is the ionizing luminosity of the source, $\phi$ is the angle between the surface normal of the wall and the line of sight from the source to the wall, $k$ is Boltzmann's constant, $T$ is the temperature in the wall, $N_s$ is the column density of the wall, $P_s$ is the pressure in the wall, $\alpha_b$ is the recombination coefficient of hydrogen, and $s$ is the distance from the shell wall to the ionizing source. Note that $N_s$ is dependent upon the initial gas density at the height of the source. Thus, by comparing the locations of the ionization front breakout point in Kompaneets models to an observationally determined location, physical parameters of a superbubble and its surrounding medium can be determined. 

Since this approach determines the location of the ionization front by assuming that the bubble walls are in ionization equilibrium with the incident ionizing photons, the only time dependence of the ionization front location comes from the evolution of the superbubble shape over time. Furthermore, since the Kompaneets model is two dimensional, the resulting ionization front structure will also be two dimensional. 

For the superbubble shape described in model A, we calculate where the ionization front lies, following the approach of \citet{Basu99}, if the initial density of the exponential atmosphere at the location of the driving source was either 0.5 or 1 cm$^{-3}$ \citep{Heiles76,Ferriere91,Brown95, Kalberla09} and the internal bubble pressure, $P / k$, is either $1 \times10^4$ or $5 \times 10^4$ K cm$^{-3}$ \citep{Burrows93, Guo95, Burrows96}. For this calculation, the gas pressure in the bubble wall is taken to be equal to the gas pressure within the bubble, although \citet{BisnovatyiKogan95} suggest that the pressure within a superbubble's wall may be twice that of the interior of the bubble. Based upon constraints from prior observations, a temperature of 8000 K is adopted for the bubble wall \citep{Basu99} and an ionizing luminosity of $4 \times 10^{49}$ s$^{-1}$ is selected \citep{Reynolds79}. A summary of the parameter range investigated is given in Table \ref{table:ion front param}. 

\begin{table}
\caption{Ionization front parameters}
\begin{center}
\begin{tabular}{ccccc}
\hline
 & $T$ & $P/k$ & $L$ & $n_0$\\
 & (K) & (K cm$^{-3}$) & (s$^{-1}$) & (cm$^{-3}$)\\
 & (1) & (2) & (3) & (4)\\
\hline
High & 8000 & $5 \times 10^4$ & $4 \times 10^{49}$ & 1\\
Low & 8000 & $1 \times 10^4$ & $4 \times 10^{49}$ & 0.5\\
\hline
\end{tabular}
\end{center}
{\it Note.} Column (1) gives the adopted temperature of the bubble wall. Columns (2)-(4), respectively, give the adopted ranges for the bubble wall's pressure, ionizing luminosity, and the initial density of the exponential atmosphere at the height of the driving source. The first row gives the highest value used and the second row gives the lowest value used.
\label{table:ion front param}
\end{table}

For a bubble with the shape of model A, ionizing photons breakout everywhere for all combinations of the above ionization parameters except for the case where the pressure, $P / k$, is $5 \times 10^4$ K cm$^{-3}$ and the initial density is 1 cm$^{-3}$. In this case, where both the pressure and the number density are high, the ionizing photons are trapped in the bottom half of the bubble. There is no combination of ionization parameters for which the ionizing photons are still trapped within the bubble wall at the distance of the Eridanus filaments. 

While \citet{Basu99} require an unusually large initial density of 10 cm$^{-3}$ to match the observed location of the ionization front in the W4 superbubble, we find that we can reproduce the observed location of the ionization front breakout in the Orion--Eridanus superbubble for an initial density and pressure that are within the plausible ranges suggested by prior observations. 

Model A overpredicts the size of the superbubble towards Barnard's Loop and this larger model size makes it easier for the bubble wall to fully trap ionizing photons. A larger density and/or pressure may be required for a model that accurately reproduces the size of Barnard's Loop. For instance, for a bubble with the shape of model T, the ionizing photons breakout of the bubble wall everywhere for the parameter ranges given in Table \ref{table:ion front param}. Conversely, the giant molecular cloud that the Orion subgroups have formed within would provide additional mass to trap ionizing photons with Barnard's Loop.
					
\section{SECONDARY PROCESSES}
\label{secondary}

	Both the elongation and orientation of the Orion--Eridanus superbubble are slightly at odds with what would naively be expected for a bubble expanding in the exponential atmosphere of the Galactic disc. In this section, we address physical processes not included in the Kompaneets model that may influence the shape of a superbubble. 
	
\subsection{Magnetic Field}
\label{mag field}	

	One potentially significant element missing from the Kompaneets model is magnetic fields, as magnetic fields hinder the expansion of a superbubble in the directions perpendicular to the magnetic fields (e.g. \citealt{Tomisaka92, Tomisaka98, Stil09}). While magnetic tension has been shown to be overpredicted in 2D simulations of expanding bubbles \citep{deAvillez05}, 3D magnetohydrodynamic (MHD) simulations of expanding superbubbles confirm that magnetic confinement can significantly alter the shape of a superbubble (e.g. \citealt{Stil09}).
	
	The local Galactic magnetic field is oriented towards a Galactic longitude of 85$^{\circ}$, parallel to the Galactic plane \citep{Rand94, Heiles96}. Assuming that the initial magnetic field direction in the vicinity of the Orion star-forming region is similarly directed, given the close proximity of the Orion star-forming region to the Sun, the magnetic field direction near the Orion--Eridanus superbubble is expected to have an inclination with respect to the plane of the sky of approximately $50^\circ$. Measurements of the direction of polarized interstellar starlight and Zeeman splitting near, but outside of the bubble, confirm that the magnetic field is parallel to the Galactic plane and inclined roughly 50$^{\circ}$ to the plane of the sky \citep{Appenzeller74, Heiles97}.
	
	Inside of the bubble, the polarization directions of starlight are not compatible with magnetic fields oriented parallel to the Galactic plane (e.g. \citealt{Appenzeller74}). A magnetic pocket model, where the magnetic field lines toward the Orion half of the bubble are pulled into an orientation more perpendicular to the Galactic plane by a Parker instability \citep{Parker66}, provides a reasonably good fit to the polarization data \citep{Appenzeller74}. \citet{Heiles97} also finds a reasonable fit with a paraboloid model, which treats the bubble as a worm/chimney structure and attributes the current magnetic field orientation to expansion motions from the bubble, rather than to a pre-existing magnetic pocket. Turbulent motions may also have altered the magnetic field direction in the gas that the bubble expanded into.  
	
	For the case of an exponential atmosphere with a magnetic field oriented perpendicular to the density gradient, as would be expected for a magnetic field parallel to the Galactic disc, \cite{Stil09} show that a superbubble will still become elongated along the axis of the density gradient, although to less of an extent than if the magnetic field were not present. Thus, it is unlikely that the Galactic magnetic field would produce a bubble oriented parallel to the Galactic plane, as in model T. The major axis of the superbubble in model T also makes an angle of $45^\circ$ with the local magnetic field, making it even more unlikely that the Galactic magnetic field would create the structure seen in model T. 
	
	The projection of the major axis of the superbubble in model A on to the Galactic plane, however, is reasonably well aligned with the magnetic field direction, with the projection being within 10$^\circ$ of the magnetic field direction. If a vertical component were added to the expected magnetic field configuration, the magnetic field would naturally channel the superbubble into an orientation similar to what is observed. Without this vertical component, the superbubble would not tilt towards the Galactic plane, as observed. 

	\citet{Stil09} also show that if the magnetic field in the Galactic plane is oriented towards the observer, Kompaneets models fit to a superbubble's shape underpredict the scaleheight of the gas by factors of 2-3 because magnetic fields create more elongated bubbles than would otherwise be expected. This could partially explain the low scaleheight derived from model  A. A preferential expansion of the bubble along the line of sight would also help explain the large ratio between the apparent depth of the bubble, 300 pc if the near side is at 180 pc and the far side is more than 500 pc distant, and the north--south width of the bubble, roughly 150 pc. While \citet{Stil09} are unable to reproduce the factor of 4-10 reduction in scaleheight required by the best-fitting models of the W4 and Orion--Eridanus superbubbles, a magnetic field oriented perpendicular to the Galactic disc would further contribute to the elongation of a superbubble (\citealt{Komljenovic99, West07}; Bailey personal communication 2008). Such an orientation of the magnetic field could be created by a Parker instability \citep{Parker66, Appenzeller74, Basu97} or by prior supernovae \citep{Heiles97, West07}. Since the magnetic field direction is 50$^{\circ}$ inclined to the plane of the sky towards the superbubble, however, the effective shortening of the width of the superbubble will be reduced compared to the best case models of \citet{Stil09}, in which the magnetic field is perfectly aligned with the line of sight. 

\subsection{Galactic Dynamics}
\label{dynamics}	
	
	Shear from Galactic differential rotation can elongate a bubble over time and will eventually decrease the minor axis of a bubble ellipsoid at a given height \citep{TenorioTagle87,Palous90}. The time-scale over which Galactic shear is expected to be effective, however, is on the order of tens to hundreds of Myr \citep{Heiles79, TenorioTagle88, Tomisaka98}, whereas the age of the oldest star-forming group in the Orion star-forming region is only 10 Myr \citep{Brown94}. 
	
	Similarly, the time-scale for the Coriolis force to become effective, 50 Myr \citep{Tomisaka98}, is too large for the Coriolis effect to be important. The Orion--Eridanus superbubble is also unlikely to be affected by a warp in the Galactic disc as the warp in the Milky Way only becomes prominent beyond a Galactocentric radius of 9 kpc \citep{Kalberla09}. Density contrasts between spiral arms and interarm regions create radial density gradients in the disc, but the density contrast between spiral arms and interarm regions is only on the order of three to five \citep{Roberts84, Vogel88}, such that it is unlikely that the arm/interarm density contrast is responsible for elongating the Orion--Eridanus superbubble. 
						
	Because of the Galactic gravitational potential, gas moving vertically from the Galactic plane will drift radially outwards from the Galactic Centre \citep{Collins02,Pidopryhora07}. While such a drift would lead to the Orion--Eridanus superbubble pointing away from the Sun, as in model A, this radial drift only becomes significant at heights of many kpc \citep{Collins02}.	

\subsection{Turbulence}
\label{turbulence}	

	Large-scale simulations of the ISM show that turbulent flows and density inhomogeneities can generate reasonably elongated superbubbles \citep{Korpi99, deAvillez05}, including bubbles that are elongated parallel to the Galactic plane \citep{deAvillez05}. Turbulent motions in the ISM not only directly influence the shape of a growing superbubble by acting on the superbubble walls, but also indirectly influence the growth of a superbubble by altering the density profile of the gas into which the superbubble will expand. We do not expect the Galactic disc to have a perfect exponential atmosphere and significant deviations from such an exponential atmosphere could channel an expanding superbubble towards particular directions.
			
\subsection{Secondary Driving Source}
\label{secondary driving source}

	There are at least four separate subgroups of stars in Orion that have formed over the last 10 Myr \citep{Brown94}. The winds and supernovae from the oldest subgroups will have altered the conditions in the ISM into which the Orion--Eridanus superbubble has been blown. Previous generations of star formation may have created low-density channels in the ISM that could have led to the elongation of the Orion--Eridanus superbubble, although it is unclear whether such previous star formation activity could induce the level of elongation observed, especially considering that all of the subgroups are on the Orion side of the bubble. \citet{Oey05} suggest that previous generations of star formation activity around the W4 superbubble could have altered the ISM into which the W4 superbubble has been blown and this may account for the low scaleheight required by the \citet{Basu99} and \citet{Baumgartner13} Kompaneets fits to the bubble.

	The ISM into which the Orion--Eridanus superbubble is expanding may also have been affected by a foreground OB subgroup. One potential source of such a subgroup is the IC 2118 cloud. IC 2118 is located 2$^\circ$ to the nort-west of Rigel and is roughly 200 pc from the Sun, placing the cloud approximately 200 pc in front of the Orion star-forming region \citep{Kun01}. IC 2118 contains on the order of 100 M$_\odot$ of molecular gas and is known to harbour tens of young stellar objects with masses up to a solar mass, which have ages on the order of a few Myr \citep{Kun01, Kun04}. Higher mass stars, and their subsequent supernovae, may have been present in IC 2118 in the past. 
	
	There are three massive stars that are all located between the Sun and the Orion--Eridanus superbubble. Betelgeuse is a 17 M$_\odot$ supergiant and its unusual proper motion has led to speculation that it is either a runaway star from IC 2118 or that it is a runaway from Orion OB1a that has undergone two velocity kicks from either supernovae or multibody interactions \citep{Bally08, Harper08}. Rigel is a massive supergiant star that is 245 pc from the Sun and externally illuminating the IC 2118 cloud \citep{Bally08}. Rigel has no significant radial velocity \citep{Kharchenko07} or proper motion \citep{Hog00} with respect to IC 2118, suggesting that Rigel may have formed near its present location. Similarly, Saiph ($\kappa$ Ori), another massive star, is located 220 pc from the Sun \citep{Bally08}. The presence of three massive stars between the Sun and Orion further suggests that an OB subgroup formed approximately 5-10 Myr ago between the Sun and Orion \citep{Bally08}. The additional 0.25 keV emission seen beyond the southern H$\alpha$ boundary of the superbubble may also be related to this OB subgroup, in line with the suggestion from \citet{Burrows93} that the extra 0.25 keV emission is generated from a star outside of the Orion star-forming region.

	The presence of additional massive stars between the Orion star-forming region and the Sun would have created an additional cavity in the ISM that, upon merging with the bubble created by the Orion star-forming region, would have created a cavity elongated towards the Sun. Since IC 2118, Betelgeuse, Rigel, and Saiph are still on the Orion side of the superbubble, it is not clear, however, whether such a merger of bubbles could explain the elongation of the Orion--Eridanus superbubble into the constellation of Eridanus. Similarly, it is unclear whether the limited star formation in IC 2118 could impact the ISM on the scales necessary to significantly alter the morphology of the Orion--Eridanus superbubble. 

An additional foreground population of 2600 stars has also been detected in front of the Orion A cloud, although these stars are still reasonably close to the Orion star-forming region with these stars being 380 pc form the Sun \citep{Alves12, Bouy14}. 

\subsection{Edge Identification}
\label{edge identification}
	The apparent elongation of the Orion--Eridanus superbubble could also be due to a misidentification of the location of the bubble wall. The H$\alpha$ features fit in Section \ref{fitting} may not trace the true exterior of the bubble. The H {\sc i} and dust (see for instance the 100 $\mu$m {\it Infrared Astronomical Satellite}, {\it IRAS}, maps of the bubble; \citealt{MivilleDeschenes05}) features lying outside of the H$\alpha$ features may trace the true walls of the bubble. Such a larger bubble would have a smaller elongation and would not require as small of a scaleheight for the exponential atmosphere. These H {\sc i} and dust features also appear to trace out a bubble that is closer to being perpendicular to the Galactic plane, at least on the plane of the sky, especially given the southern extension of H {\sc i} filaments towards Arc C. It is possible that some of the observed H$\alpha$ emission could be coming from clouds within the bubble that were not fully swept into the wall of the bubble or could be coming from material injected into the bubble from the Orion star-forming region. The extension of the 0.25 keV X-ray emission to the south of the H$\alpha$ boundary would also agree with the larger bubble shape suggested by the H {\sc i} data. Since the soft 0.75 keV X-ray enhancement is relatively well confined by the H$\alpha$ filaments, however, we consider it unlikely that the bubble walls extend significantly past the apparent H$\alpha$ boundary of the superbubble.
		
\section{CONCLUSIONS}
\label{conclusions}

The Orion star-forming region is the closest high-mass star-forming region currently forming stars and it has blown a large 20$^{\circ}$ x 45$^{\circ}$ superbubble into the ISM. We fit Kompaneets models, models of bubbles driven by continuous energy sources expanding into exponential atmospheres, to the morphology of the Orion--Eridanus superbubble, as traced in H$\alpha$. We are unable to find a reasonable model fit where the superbubble is oriented with the Eridanus side closer to the Sun and the near side of the superbubble is approximately 180 pc away. A model (model A) where the Eridanus side of the superbubble is oriented away from the Sun and the back side is approximately 500 pc distant, however, provides a marginal fit to the data. This model is slightly inclined from the normal to the Galactic plane, by 35$^\circ$, and has a slightly unusually small scaleheight of 40 pc for the Galactic disc's exponential atmosphere, although this small scaleheight is consistent with the scaleheight of 25 pc found for the W4 superbubble \citep{Basu99, Baumgartner13}. We propose that model A should be adopted as a general framework for understanding the Orion--Eridanus superbubble, with some secondary process required to account for the elongation of the superbubble and the offset between the superbubble's major axis and the normal to the Galactic plane. Possibilities for this second process include magnetic fields, turbulent flows, Galactic dynamics, and additional driving sources. In particular, the combination of model A with magnetic fields and/or a secondary driving source provides a reasonably compelling solution for the shape of the Orion--Eridanus superbubble. 

The Orion--Eridanus superbubble provides a further example of a superbubble that is unusually elongated (e.g. \citealt{Basu99}) and oriented away from the normal to the Galactic plane  (e.g. \citealt{Suad14}), such that physics beyond that included in the Kompaneets model are required to fully explain the observed shape of the superbubble. Since superbubbles play an important role in Galactic ecology by funnelling hot gas out of the Galactic plane and into the Galactic halo, it is vital that this secondary process be identified and understood. An increase in the typical elongation of a superbubble will promote the transport of hot gas into the halo while a shift of the major axis of a superbubble towards the Galactic plane will tend to trap hot gas closer to the Galactic plane. The Orion--Eridanus superbubble presents a prime candidate for further study into this secondary process, given the superbubble's nearby location. The orientation of the Orion--Eridanus superbubble, towards or away from the Sun, also provides an observable test on the importance of this secondary process.
	
	Based upon prior observations, we suggest that an ionization front breaks out of the Orion--Eridanus superbubble wall just past the western edges of Barnard's Loop. We fit multiple ionization front models to our best-fitting Kompaneets model and find that we can reproduce the observed location of the ionization front breakout if the interior bubble pressure, $P / k$, is $5 \times 10^4$ K cm$^{-3}$ and the initial density of the exponential atmosphere at the height of the Orion star-forming region is 1 cm$^{-3}$. Based upon the shape of our best-fitting model, we suggest that LDN 1551, LDN 1558, and LDN 1563 lie outside of the superbubble's wall, although these clouds should still be sculpted by ionizing photons from the Orion star-forming region that penetrate the superbubble's wall past the edges of Barnard's Loop. 
		 
\section*{ACKNOWLEDGEMENTS}
	We would like to thank our anonymous referee for many useful changes to this paper. We would like to heartily thank Dr Basu for providing some of his code to calculate the surface densities of Kompaneets bubbles, as well as for general advice on dealing with superbubbles. We would also like to thank Dr Vaidya and Dr Caselli for useful discussions and comments. AP was partially supported by the Natural Sciences and Engineering Research Council of Canada graduate scholarship program. DJ acknowledges support from a Natural Sciences and Engineering Research Council (NSERC) Discovery Grant. This research has made use of the Smithsonian Astrophysical Observatory (SAO)/National Aeronautics and Space Administration's (NASA's) Astrophysics Data System (ADS). The Wisconsin H-Alpha Mapper (WHAM) is funded by the National Science Foundation. This research has made use of the SIMBAD data base, operated at CDS, Strasbourg, France. 
\bibliographystyle{mn2e}
\bibliography{ponbib}{}

\begin{thebibliography}{}
 \providecommand{\href}[2]{#2}

\bibitem[\protect\citeauthoryear{{Alves} \& {Bouy}}{{Alves} \&
  {Bouy}}{2012}]{Alves12}
{Alves} J.,  {Bouy} H.,  2012, \aap, 547, A97

\bibitem[\protect\citeauthoryear{{Appenzeller}}{{Appenzeller}}{1974}]{Appenzeller74}
{Appenzeller} I.,  1974, \aap, 36, 99

\bibitem[\protect\citeauthoryear{{Arnal}, {Bajaja}, {Larrarte}, {Morras} \&
  {P{\"o}ppel}}{{Arnal} et~al.}{2000}]{Arnal00}
{Arnal} E.~M.,  {Bajaja} E.,  {Larrarte} J.~J.,  {Morras} R.,    {P{\"o}ppel}
  W.~G.~L.,  2000, \aaps, 142, 35

\bibitem[\protect\citeauthoryear{{Bagetakos}, {Brinks}, {Walter}, {de Blok},
  {Usero}, {Leroy}, {Rich} \& {Kennicutt} Jr.}{{Bagetakos}
  et~al.}{2011}]{Bagetakos11}
{Bagetakos} I.,  {Brinks} E.,  {Walter} F.,  {de Blok} W.~J.~G.,  {Usero} A.,
  {Leroy} A.~K.,  {Rich} J.~W.,    {Kennicutt} Jr. R.~C.,  2011, \aj, 141, 23

\bibitem[\protect\citeauthoryear{{Bajaja}, {Arnal}, {Larrarte}, {Morras},
  {P{\"o}ppel} \& {Kalberla}}{{Bajaja} et~al.}{2005}]{Bajaja05}
{Bajaja} E.,  {Arnal} E.~M.,  {Larrarte} J.~J.,  {Morras} R.,  {P{\"o}ppel}
  W.~G.~L.,    {Kalberla} P.~M.~W.,  2005, \aap, 440, 767

\bibitem[\protect\citeauthoryear{{Bally}}{{Bally}}{2001}]{Bally01}
{Bally} J.,  2001, in {Woodward} C.~E.,  {Bicay} M.~D.,   {Shull} J.~M.,  eds,
  Astronomical Society of the Pacific Conference Series Vol. 231, Tetons 4:
  Galactic Structure, Stars and the Interstellar Medium. {San Francisco, CA:
  ASP}, p.~204

\bibitem[\protect\citeauthoryear{{Bally}}{{Bally}}{2008}]{Bally08}
{Bally} J.,  2008, in {Reipurth} B.,  ed., Handbook of Star Forming Regions,
  Volume I. {San Francisco, CA: ASP Monograph Publications}, p.~459

\bibitem[\protect\citeauthoryear{{Barnard}}{{Barnard}}{1894}]{Barnard1894}
{Barnard} E.~E.,  1894, Popular Astronomy, 2, 151

\bibitem[\protect\citeauthoryear{{Basu}, {Mouschovias} \&
  {Paleologou}}{\protect\mniiiauthor{Basu97}{{Basu}, {Mouschovias} \&
  {Paleologou}}{{Basu} et~al.}}{1997}]{Basu97}
{Basu} S.,  {Mouschovias} T.~C.,    {Paleologou} E.~V.,  1997, \apjl, 480, L55

\bibitem[\protect\citeauthoryear{{Basu}, {Johnstone} \&
  {Martin}}{\protect\mniiiauthor{Basu99}{{Basu}, {Johnstone} \&
  {Martin}}{{Basu} et~al.}}{1999}]{Basu99}
{Basu} S.,  {Johnstone} D.,    {Martin} P.~G.,  1999, \apj, 516, 843

\bibitem[\protect\citeauthoryear{{Baumgartner} \&
  {Breitschwerdt}}{{Baumgartner} \& {Breitschwerdt}}{2013}]{Baumgartner13}
{Baumgartner} V.,  {Breitschwerdt} D.,  2013, \aap, 557, A140

\bibitem[\protect\citeauthoryear{{Bisnovatyi-Kogan} \&
  {Silich}}{{Bisnovatyi-Kogan} \& {Silich}}{1995}]{BisnovatyiKogan95}
{Bisnovatyi-Kogan} G.~S.,  {Silich} S.~A.,  1995, Reviews of Modern Physics,
  67, 661

\bibitem[\protect\citeauthoryear{{Bisnovatyi-Kogan}, {Blinnikov} \&
  {Silich}}{\protect\mniiiauthor{BisnovatyiKogan89}{{Bisnovatyi-Kogan},
  {Blinnikov} \& {Silich}}{{Bisnovatyi-Kogan}
  et~al.}}{1989}]{BisnovatyiKogan89}
{Bisnovatyi-Kogan} G.~S.,  {Blinnikov} S.~I.,    {Silich} S.~A.,  1989, \apss,
  154, 229

\bibitem[\protect\citeauthoryear{{Boumis}, {Dickinson}, {Meaburn}, {Goudis},
  {Christopoulou}, {L{\'o}pez}, {Bryce} \& {Redman}}{{Boumis}
  et~al.}{2001}]{Boumis01}
{Boumis} P.,  {Dickinson} C.,  {Meaburn} J.,  {Goudis} C.~D.,  {Christopoulou}
  P.~E.,  {L{\'o}pez} J.~A.,  {Bryce} M.,    {Redman} M.~P.,  2001, \mnras,
  320, 61

\bibitem[\protect\citeauthoryear{{Bouy}, {Alves}, {Bertin}, {Sarro} \&
  {Barrado}}{{Bouy} et~al.}{2014}]{Bouy14}
{Bouy} H.,  {Alves} J.,  {Bertin} E.,  {Sarro} L.~M.,    {Barrado} D.,  2014,
  \aap, 564, A29

\bibitem[\protect\citeauthoryear{{Brown}, {de Geus} \& {de
  Zeeuw}}{\protect\mniiiauthor{Brown94}{{Brown}, {de Geus} \& {de
  Zeeuw}}{{Brown} et~al.}}{1994}]{Brown94}
{Brown} A.~G.~A.,  {de Geus} E.~J.,    {de Zeeuw} P.~T.,  1994, \aap, 289, 101

\bibitem[\protect\citeauthoryear{{Brown}, {Hartmann} \&
  {Burton}}{\protect\mniiiauthor{Brown95}{{Brown}, {Hartmann} \&
  {Burton}}{{Brown} et~al.}}{1995}]{Brown95}
{Brown} A.~G.~A.,  {Hartmann} D.,    {Burton} W.~B.,  1995, \aap, 300, 903

\bibitem[\protect\citeauthoryear{{Burrows} \& {Zhiyu}}{{Burrows} \&
  {Zhiyu}}{1996}]{Burrows96}
{Burrows} D.~N.,  {Zhiyu} G.,  1996, in {Zimmermann} H.~U.,  {Tr{\"u}mper} J.,
   {Yorke} H.,  eds, Roentgenstrahlung from the Universe. {MPE Rep. 263;
  Garching: MPE}, p.~221

\bibitem[\protect\citeauthoryear{{Burrows}, {Singh}, {Nousek}, {Garmire} \&
  {Good}}{{Burrows} et~al.}{1993}]{Burrows93}
{Burrows} D.~N.,  {Singh} K.~P.,  {Nousek} J.~A.,  {Garmire} G.~P.,    {Good}
  J.,  1993, \apj, 406, 97

\bibitem[\protect\citeauthoryear{{Churchwell} et~al.,}{{Churchwell}
  et~al.}{2006}]{Churchwell06}
{Churchwell} E.  et~al., 2006, \apj, 649, 759

\bibitem[\protect\citeauthoryear{{Churchwell} et~al.,}{{Churchwell}
  et~al.}{2007}]{Churchwell07}
{Churchwell} E.  et~al., 2007, \apj, 670, 428

\bibitem[\protect\citeauthoryear{{Collins}, {Benjamin} \&
  {Rand}}{\protect\mniiiauthor{Collins02}{{Collins}, {Benjamin} \&
  {Rand}}{{Collins} et~al.}}{2002}]{Collins02}
{Collins} J.~A.,  {Benjamin} R.~A.,    {Rand} R.~J.,  2002, \apj, 578, 98

\bibitem[\protect\citeauthoryear{{Davidsen}, {Shulman}, {Fritz}, {Meekins},
  {Henry} \& {Friedman}}{{Davidsen} et~al.}{1972}]{Davidsen72}
{Davidsen} A.,  {Shulman} S.,  {Fritz} G.,  {Meekins} J.~F.,  {Henry} R.~C.,
  {Friedman} H.,  1972, \apj, 177, 629

\bibitem[\protect\citeauthoryear{{Di Cicco} \& {Walker}}{{Di Cicco} \&
  {Walker}}{2009}]{DiCicco09}
{Di Cicco} D.,  {Walker} S.,  2009, \skytel, 117, 66

\bibitem[\protect\citeauthoryear{{Ehlerov{\'a}} \& {Palou{\v
  s}}}{{Ehlerov{\'a}} \& {Palou{\v s}}}{2005}]{Ehlerova05}
{Ehlerov{\'a}} S.,  {Palou{\v s}} J.,  2005, \aap, 437, 101

\bibitem[\protect\citeauthoryear{{Ehlerov{\'a}} \& {Palou{\v
  s}}}{{Ehlerov{\'a}} \& {Palou{\v s}}}{2013}]{Ehlerova13}
{Ehlerov{\'a}} S.,  {Palou{\v s}} J.,  2013, \aap, 550, A23

\bibitem[\protect\citeauthoryear{{Elliott}}{{Elliott}}{1973}]{Elliott73}
{Elliott} K.~H.,  1973, \aap, 26, 279

\bibitem[\protect\citeauthoryear{{Elliott} \& {Meaburn}}{{Elliott} \&
  {Meaburn}}{1970}]{Elliott70}
{Elliott} K.~H.,  {Meaburn} J.,  1970, \apss, 7, 252

\bibitem[\protect\citeauthoryear{{Ferriere}, {Mac Low} \&
  {Zweibel}}{\protect\mniiiauthor{Ferriere91}{{Ferriere}, {Mac Low} \&
  {Zweibel}}{{Ferriere} et~al.}}{1991}]{Ferriere91}
{Ferriere} K.~M.,  {Mac Low} M.-M.,    {Zweibel} E.~G.,  1991, \apj, 375, 239

\bibitem[\protect\citeauthoryear{{Fried}, {Nousek}, {Sanders} \&
  {Kraushaar}}{{Fried} et~al.}{1980}]{Fried80}
{Fried} P.~M.,  {Nousek} J.~A.,  {Sanders} W.~T.,    {Kraushaar} W.~L.,  1980,
  \apj, 242, 987

\bibitem[\protect\citeauthoryear{{Frisch}, {Sembach} \&
  {York}}{\protect\mniiiauthor{Frisch90}{{Frisch}, {Sembach} \&
  {York}}{{Frisch} et~al.}}{1990}]{Frisch90}
{Frisch} P.~C.,  {Sembach} K.,    {York} D.~G.,  1990, \apj, 364, 540

\bibitem[\protect\citeauthoryear{{Garmire}, {Nousek}, {Apparao}, {Burrows},
  {Fink} \& {Kraft}}{{Garmire} et~al.}{1992}]{Garmire92}
{Garmire} G.~P.,  {Nousek} J.~A.,  {Apparao} K.~M.~V.,  {Burrows} D.~N.,
  {Fink} R.~L.,    {Kraft} R.~P.,  1992, \apj, 399, 694

\bibitem[\protect\citeauthoryear{{Green}}{{Green}}{1991}]{Green91}
{Green} D.~A.,  1991, \mnras, 253, 350

\bibitem[\protect\citeauthoryear{{Guo}, {Burrows}, {Sanders}, {Snowden} \&
  {Penprase}}{{Guo} et~al.}{1995}]{Guo95}
{Guo} Z.,  {Burrows} D.~N.,  {Sanders} W.~T.,  {Snowden} S.~L.,    {Penprase}
  B.~E.,  1995, \apj, 453, 256

\bibitem[\protect\citeauthoryear{{Haffner}, {Reynolds}, {Tufte}, {Madsen},
  {Jaehnig} \& {Percival}}{{Haffner} et~al.}{2003}]{Haffner03}
{Haffner} L.~M.,  {Reynolds} R.~J.,  {Tufte} S.~L.,  {Madsen} G.~J.,  {Jaehnig}
  K.~P.,    {Percival} J.~W.,  2003, \apjs, 149, 405

\bibitem[\protect\citeauthoryear{{Harper}, {Brown} \&
  {Guinan}}{\protect\mniiiauthor{Harper08}{{Harper}, {Brown} \&
  {Guinan}}{{Harper} et~al.}}{2008}]{Harper08}
{Harper} G.~M.,  {Brown} A.,    {Guinan} E.~F.,  2008, \aj, 135, 1430

\bibitem[\protect\citeauthoryear{{Hartmann} \& {Burton}}{{Hartmann} \&
  {Burton}}{1997}]{Hartmann97}
{Hartmann} D.,  {Burton} W.~B.,  1997, {Atlas of Galactic Neutral Hydrogen}.
{Cambridge: Cambridge Univ. Press}

\bibitem[\protect\citeauthoryear{{Heiles}}{{Heiles}}{1976}]{Heiles76}
{Heiles} C.,  1976, \apjl, 208, L137

\bibitem[\protect\citeauthoryear{{Heiles}}{{Heiles}}{1979}]{Heiles79}
{Heiles} C.,  1979, \apj, 229, 533

\bibitem[\protect\citeauthoryear{{Heiles}}{{Heiles}}{1987}]{Heiles87}
{Heiles} C.,  1987, \apj, 315, 555

\bibitem[\protect\citeauthoryear{{Heiles}}{{Heiles}}{1996}]{Heiles96}
{Heiles} C.,  1996, \apj, 462, 316

\bibitem[\protect\citeauthoryear{{Heiles}}{{Heiles}}{1997}]{Heiles97}
{Heiles} C.,  1997, \apjs, 111, 245

\bibitem[\protect\citeauthoryear{{Heiles} \& {Habing}}{{Heiles} \&
  {Habing}}{1974}]{Heiles74Habing}
{Heiles} C.,  {Habing} H.~J.,  1974, \aaps, 14, 1

\bibitem[\protect\citeauthoryear{{Heiles}, {Haffner} \&
  {Reynolds}}{\protect\mniiiauthor{Heiles99}{{Heiles}, {Haffner} \&
  {Reynolds}}{{Heiles} et~al.}}{1999}]{Heiles99}
{Heiles} C.,  {Haffner} L.~M.,    {Reynolds} R.~J.,  1999, in {Taylor} A.~R.,
  {Landecker} T.~L.,   {Joncas} G.,  eds,  Astronomical Society of the Pacific
  Conference Series Vol. 168, New Perspectives on the Interstellar Medium. {San
  Francisco, CA: ASP}, p.~211

\bibitem[\protect\citeauthoryear{{Heyer}, {Brunt}, {Snell}, {Howe}, {Schloerb}
  \& {Carpenter}}{{Heyer} et~al.}{1998}]{Heyer98}
{Heyer} M.~H.,  {Brunt} C.,  {Snell} R.~L.,  {Howe} J.~E.,  {Schloerb} F.~P.,
   {Carpenter} J.~M.,  1998, \apjs, 115, 241

\bibitem[\protect\citeauthoryear{{Hirota} et~al.,}{{Hirota}
  et~al.}{2007}]{Hirota07}
{Hirota} T.  et~al., 2007, \pasj, 59, 897

\bibitem[\protect\citeauthoryear{{H{\o}g} et~al.,}{{H{\o}g}
  et~al.}{2000}]{Hog00}
{H{\o}g} E.  et~al., 2000, \aap, 355, L27

\bibitem[\protect\citeauthoryear{{Isobe}}{{Isobe}}{1973}]{Isobe73}
{Isobe} S.,  1973, in {Greenberg} J.~M.,  {van de Hulst} H.~C.,  eds,  IAU
  Symposium Vol. 52, Interstellar Dust and Related Topics. {Dordrecht: Reidel},
  p.~433

\bibitem[\protect\citeauthoryear{{Johnson}}{{Johnson}}{1978}]{Johnson78}
{Johnson} P.~G.,  1978, \mnras, 184, 727

\bibitem[\protect\citeauthoryear{{Kalberla} \& {Kerp}}{{Kalberla} \&
  {Kerp}}{2009}]{Kalberla09}
{Kalberla} P.~M.~W.,  {Kerp} J.,  2009, \araa, 47, 27

\bibitem[\protect\citeauthoryear{{Kalberla}, {Burton}, {Hartmann}, {Arnal},
  {Bajaja}, {Morras} \& {P{\"o}ppel}}{{Kalberla} et~al.}{2005}]{Kalberla05}
{Kalberla} P.~M.~W.,  {Burton} W.~B.,  {Hartmann} D.,  {Arnal} E.~M.,  {Bajaja}
  E.,  {Morras} R.,    {P{\"o}ppel} W.~G.~L.,  2005, \aap, 440, 775

\bibitem[\protect\citeauthoryear{{Kharchenko}, {Scholz}, {Piskunov},
  {R{\"o}ser} \& {Schilbach}}{{Kharchenko} et~al.}{2007}]{Kharchenko07}
{Kharchenko} N.~V.,  {Scholz} R.-D.,  {Piskunov} A.~E.,  {R{\"o}ser} S.,
  {Schilbach} E.,  2007, Astronomische Nachrichten, 328, 889

\bibitem[\protect\citeauthoryear{{Komljenovic}, {Basu} \&
  {Johnstone}}{\protect\mniiiauthor{Komljenovic99}{{Komljenovic}, {Basu} \&
  {Johnstone}}{{Komljenovic} et~al.}}{1999}]{Komljenovic99}
{Komljenovic} P.~T.,  {Basu} S.,    {Johnstone} D.,  1999, in {Taylor} A.~R.,
  {Landecker} T.~L.,   {Joncas} G.,  eds,  Astronomical Society of the Pacific
  Conference Series Vol. 168, New Perspectives on the Interstellar Medium. {San
  Francisco, CA: ASP}, p.~299

\bibitem[\protect\citeauthoryear{{Kompaneets}}{{Kompaneets}}{1960}]{Kompaneets60}
{Kompaneets} A.~S.,  1960, Sov. Phys. Dokl., 5, 46

\bibitem[\protect\citeauthoryear{{Korpi}, {Brandenburg}, {Shukurov} \&
  {Tuominen}}{{Korpi} et~al.}{1999}]{Korpi99}
{Korpi} M.~J.,  {Brandenburg} A.,  {Shukurov} A.,    {Tuominen} I.,  1999,
  \aap, 350, 230

\bibitem[\protect\citeauthoryear{{Kun}, {Aoyama}, {Yoshikawa}, {Kawamura},
  {Yonekura}, {Onishi} \& {Fukui}}{{Kun} et~al.}{2001}]{Kun01}
{Kun} M.,  {Aoyama} H.,  {Yoshikawa} N.,  {Kawamura} A.,  {Yonekura} Y.,
  {Onishi} T.,    {Fukui} Y.,  2001, \pasj, 53, 1063

\bibitem[\protect\citeauthoryear{{Kun}, {Prusti}, {Nikoli{\'c}}, {Johansson} \&
  {Walton}}{{Kun} et~al.}{2004}]{Kun04}
{Kun} M.,  {Prusti} T.,  {Nikoli{\'c}} S.,  {Johansson} L.~E.~B.,    {Walton}
  N.~A.,  2004, \aap, 418, 89

\bibitem[\protect\citeauthoryear{{Lee} \& {Chen}}{{Lee} \&
  {Chen}}{2009}]{Lee09}
{Lee} H.-T.,  {Chen} W.~P.,  2009, \apj, 694, 1423

\bibitem[\protect\citeauthoryear{{Lockman}}{{Lockman}}{1984}]{Lockman84}
{Lockman} F.~J.,  1984, \apj, 283, 90

\bibitem[\protect\citeauthoryear{{Long}, {Patterson}, {Moore} \&
  {Garmire}}{{Long} et~al.}{1977}]{Long77}
{Long} K.~S.,  {Patterson} J.~R.,  {Moore} W.~E.,    {Garmire} G.~P.,  1977,
  \apj, 212, 427

\bibitem[\protect\citeauthoryear{{Lopez}, {Krumholz}, {Bolatto}, {Prochaska},
  {Ramirez-Ruiz} \& {Castro}}{{Lopez} et~al.}{2013}]{Lopez13}
{Lopez} L.~A.,  {Krumholz} M.~R.,  {Bolatto} A.~D.,  {Prochaska} J.~X.,
  {Ramirez-Ruiz} E.,    {Castro} D.,  2013, preprint (arXiv:1309.5421)

\bibitem[\protect\citeauthoryear{{Mac Low} \& {McCray}}{{Mac Low} \&
  {McCray}}{1988}]{MacLow88}
{Mac Low} M.-M.,  {McCray} R.,  1988, \apj, 324, 776

\bibitem[\protect\citeauthoryear{{Mac Low}, {McCray} \&
  {Norman}}{\protect\mniiiauthor{MacLow89}{{Mac Low}, {McCray} \&
  {Norman}}{{Mac Low} et~al.}}{1989}]{MacLow89}
{Mac Low} M.-M.,  {McCray} R.,    {Norman} M.~L.,  1989, \apj, 337, 141

\bibitem[\protect\citeauthoryear{{Maciejewski} \& {Cox}}{{Maciejewski} \&
  {Cox}}{1999}]{Maciejewski99}
{Maciejewski} W.,  {Cox} D.~P.,  1999, \apj, 511, 792

\bibitem[\protect\citeauthoryear{{Marshall} \& {Clark}}{{Marshall} \&
  {Clark}}{1984}]{Marshall84}
{Marshall} F.~J.,  {Clark} G.~W.,  1984, \apj, 287, 633

\bibitem[\protect\citeauthoryear{{McCammon}, {Burrows}, {Sanders} \&
  {Kraushaar}}{{McCammon} et~al.}{1983}]{McCammon83}
{McCammon} D.,  {Burrows} D.~N.,  {Sanders} W.~T.,    {Kraushaar} W.~L.,  1983,
  \apj, 269, 107

\bibitem[\protect\citeauthoryear{{McCray} \& {Kafatos}}{{McCray} \&
  {Kafatos}}{1987}]{McCray87}
{McCray} R.,  {Kafatos} M.,  1987, \apj, 317, 190

\bibitem[\protect\citeauthoryear{{McKee}, {van Buren} \&
  {Lazareff}}{\protect\mniiiauthor{McKee84}{{McKee}, {van Buren} \&
  {Lazareff}}{{McKee} et~al.}}{1984}]{McKee84}
{McKee} C.~F.,  {van Buren} D.,    {Lazareff} B.,  1984, \apjl, 278, L115

\bibitem[\protect\citeauthoryear{{Meaburn}}{{Meaburn}}{1965}]{Meaburn65}
{Meaburn} J.,  1965, \nat, 208, 575

\bibitem[\protect\citeauthoryear{{Meaburn}}{{Meaburn}}{1967}]{Meaburn67}
{Meaburn} J.,  1967, \zap, 65, 93

\bibitem[\protect\citeauthoryear{{Menon}}{{Menon}}{1957}]{Menon57}
{Menon} T.~K.,  1957, in {van de Hulst} H.~C.,  ed.,  IAU Symposium Vol. 4,
  Radio astronomy. {Cambridge: Cambridge University Press}, p.~56

\bibitem[\protect\citeauthoryear{{Menten}, {Reid}, {Forbrich} \&
  {Brunthaler}}{{Menten} et~al.}{2007}]{Menten07}
{Menten} K.~M.,  {Reid} M.~J.,  {Forbrich} J.,    {Brunthaler} A.,  2007, \aap,
  474, 515

\bibitem[\protect\citeauthoryear{{Miville-Desch{\^e}nes} \&
  {Lagache}}{{Miville-Desch{\^e}nes} \& {Lagache}}{2005}]{MivilleDeschenes05}
{Miville-Desch{\^e}nes} M.-A.,  {Lagache} G.,  2005, \apjs, 157, 302

\bibitem[\protect\citeauthoryear{{Moriarty-Schieven}, {Johnstone}, {Bally} \&
  {Jenness}}{{Moriarty-Schieven} et~al.}{2006}]{MoriartySchieven06}
{Moriarty-Schieven} G.~H.,  {Johnstone} D.,  {Bally} J.,    {Jenness} T.,
  2006, \apj, 645, 357

\bibitem[\protect\citeauthoryear{{Naranan}, {Shulman}, {Friedman} \&
  {Fritz}}{{Naranan} et~al.}{1976}]{Naranan76}
{Naranan} S.,  {Shulman} S.,  {Friedman} H.,    {Fritz} G.,  1976, \apj, 208,
  718

\bibitem[\protect\citeauthoryear{{Nousek}, {Fried}, {Sanders} \&
  {Kraushaar}}{{Nousek} et~al.}{1982}]{Nousek82}
{Nousek} J.~A.,  {Fried} P.~M.,  {Sanders} W.~T.,    {Kraushaar} W.~L.,  1982,
  \apj, 258, 83

\bibitem[\protect\citeauthoryear{{Oey}, {Watson}, {Kern} \& {Walth}}{{Oey}
  et~al.}{2005}]{Oey05}
{Oey} M.~S.,  {Watson} A.~M.,  {Kern} K.,    {Walth} G.~L.,  2005, \aj, 129,
  393

\bibitem[\protect\citeauthoryear{{Palous}, {Franco} \&
  {Tenorio-Tagle}}{\protect\mniiiauthor{Palous90}{{Palous}, {Franco} \&
  {Tenorio-Tagle}}{{Palous} et~al.}}{1990}]{Palous90}
{Palous} J.,  {Franco} J.,    {Tenorio-Tagle} G.,  1990, \aap, 227, 175

\bibitem[\protect\citeauthoryear{{Parker}}{{Parker}}{1966}]{Parker66}
{Parker} E.~N.,  1966, \apj, 145, 811

\bibitem[\protect\citeauthoryear{{Pickering}}{{Pickering}}{1890}]{Pickering1890}
{Pickering} W.~H.,  1890, The Sidereal Messenger, 9, 2

\bibitem[\protect\citeauthoryear{{Pidopryhora}, {Lockman} \&
  {Shields}}{\protect\mniiiauthor{Pidopryhora07}{{Pidopryhora}, {Lockman} \&
  {Shields}}{{Pidopryhora} et~al.}}{2007}]{Pidopryhora07}
{Pidopryhora} Y.,  {Lockman} F.~J.,    {Shields} J.~C.,  2007, \apj, 656, 928

\bibitem[\protect\citeauthoryear{{Pon}}{{Pon}}{2013}]{Pon13Thesis}
{Pon} A.,  2013, PhD thesis, University of Victoria

\bibitem[\protect\citeauthoryear{{Pon}, {Johnstone}, {Bally} \& {Heiles}}{{Pon}
  et~al.}{2014}]{Pon14a}
{Pon} A.,  {Johnstone} D.,  {Bally} J.,    {Heiles} C.,  2014, \mnras, 441,
  1095

\bibitem[\protect\citeauthoryear{{Rand} \& {Lyne}}{{Rand} \&
  {Lyne}}{1994}]{Rand94}
{Rand} R.~J.,  {Lyne} A.~G.,  1994, \mnras, 268, 497

\bibitem[\protect\citeauthoryear{{Reynolds}}{{Reynolds}}{1989}]{Reynolds89}
{Reynolds} R.~J.,  1989, \apjl, 339, L29

\bibitem[\protect\citeauthoryear{{Reynolds} \& {Ogden}}{{Reynolds} \&
  {Ogden}}{1979}]{Reynolds79}
{Reynolds} R.~J.,  {Ogden} P.~M.,  1979, \apj, 229, 942

\bibitem[\protect\citeauthoryear{{Reynolds}, {Roesler} \&
  {Scherb}}{\protect\mniiiauthor{Reynolds74}{{Reynolds}, {Roesler} \&
  {Scherb}}{{Reynolds} et~al.}}{1974}]{Reynolds74}
{Reynolds} R.~J.,  {Roesler} F.~L.,    {Scherb} F.,  1974, \apjl, 192, L53

\bibitem[\protect\citeauthoryear{{Roberts} Jr. \& {Hausman}}{{Roberts} \&
  {Hausman}}{1984}]{Roberts84}
{Roberts} Jr. W.~W.,  {Hausman} M.~A.,  1984, \apj, 277, 744

\bibitem[\protect\citeauthoryear{{Ryu} et~al.,}{{Ryu} et~al.}{2006}]{Ryu06}
{Ryu} K.  et~al., 2006, \apjl, 644, L185

\bibitem[\protect\citeauthoryear{{Sandstrom}, {Peek}, {Bower}, {Bolatto} \&
  {Plambeck}}{{Sandstrom} et~al.}{2007}]{Sandstrom07}
{Sandstrom} K.~M.,  {Peek} J.~E.~G.,  {Bower} G.~C.,  {Bolatto} A.~D.,
  {Plambeck} R.~L.,  2007, \apj, 667, 1161

\bibitem[\protect\citeauthoryear{{Savage}, {Sembach} \&
  {Lu}}{\protect\mniiiauthor{Savage97}{{Savage}, {Sembach} \& {Lu}}{{Savage}
  et~al.}}{1997}]{Savage97}
{Savage} B.~D.,  {Sembach} K.~R.,    {Lu} L.,  1997, \aj, 113, 2158

\bibitem[\protect\citeauthoryear{{Schiano}}{{Schiano}}{1985}]{Schiano85}
{Schiano} A.~V.~R.,  1985, \apj, 299, 24

\bibitem[\protect\citeauthoryear{{Shields}}{{Shields}}{1990}]{Shields90}
{Shields} G.~A.,  1990, \araa, 28, 525

\bibitem[\protect\citeauthoryear{{Shull} \& {Saken}}{{Shull} \&
  {Saken}}{1995}]{Shull95}
{Shull} J.~M.,  {Saken} J.~M.,  1995, \apj, 444, 663

\bibitem[\protect\citeauthoryear{{Singh}, {Manchanda}, {Narana} \&
  {Sreekantan}}{{Singh} et~al.}{1982}]{Singh82}
{Singh} K.~P.,  {Manchanda} R.~K.,  {Narana} S.,    {Sreekantan} B.~V.,  1982,
  \aplett, 23, 47

\bibitem[\protect\citeauthoryear{{Sivan}}{{Sivan}}{1974}]{Sivan74}
{Sivan} J.~P.,  1974, \aaps, 16, 163

\bibitem[\protect\citeauthoryear{{Snowden}, {Burrows}, {Sanders}, {Aschenbach}
  \& {Pfeffermann}}{{Snowden} et~al.}{1995}]{Snowden95Burrows}
{Snowden} S.~L.,  {Burrows} D.~N.,  {Sanders} W.~T.,  {Aschenbach} B.,
  {Pfeffermann} E.,  1995, \apj, 439, 399

\bibitem[\protect\citeauthoryear{{Snowden} et~al.,}{{Snowden}
  et~al.}{1997}]{Snowden97}
{Snowden} S.~L.  et~al., 1997, \apj, 485, 125

\bibitem[\protect\citeauthoryear{{Staveley-Smith}, {Sault}, {Hatzidimitriou},
  {Kesteven} \& {McConnell}}{{Staveley-Smith} et~al.}{1997}]{StaveleySmith97}
{Staveley-Smith} L.,  {Sault} R.~J.,  {Hatzidimitriou} D.,  {Kesteven} M.~J.,
   {McConnell} D.,  1997, \mnras, 289, 225

\bibitem[\protect\citeauthoryear{{Stil}, {Wityk}, {Ouyed} \& {Taylor}}{{Stil}
  et~al.}{2009}]{Stil09}
{Stil} J.,  {Wityk} N.,  {Ouyed} R.,    {Taylor} A.~R.,  2009, \apj, 701, 330

\bibitem[\protect\citeauthoryear{{Suad}, {Caiafa}, {Arnal} \&
  {Cichowolski}}{{Suad} et~al.}{2014}]{Suad14}
{Suad} L.~A.,  {Caiafa} C.~F.,  {Arnal} E.~M.,    {Cichowolski} S.,  2014,
  \aap, 564, A116

\bibitem[\protect\citeauthoryear{{Tenorio-Tagle} \&
  {Bodenheimer}}{{Tenorio-Tagle} \& {Bodenheimer}}{1988}]{TenorioTagle88}
{Tenorio-Tagle} G.,  {Bodenheimer} P.,  1988, \araa, 26, 145

\bibitem[\protect\citeauthoryear{{Tenorio-Tagle} \& {Palous}}{{Tenorio-Tagle}
  \& {Palous}}{1987}]{TenorioTagle87}
{Tenorio-Tagle} G.,  {Palous} J.,  1987, \aap, 186, 287

\bibitem[\protect\citeauthoryear{{Tenorio-Tagle}, {Rozyczka} \&
  {Bodenheimer}}{\protect\mniiiauthor{TenorioTagle90}{{Tenorio-Tagle},
  {Rozyczka} \& {Bodenheimer}}{{Tenorio-Tagle} et~al.}}{1990}]{TenorioTagle90}
{Tenorio-Tagle} G.,  {Rozyczka} M.,    {Bodenheimer} P.,  1990, \aap, 237, 207

\bibitem[\protect\citeauthoryear{{Tomisaka}}{{Tomisaka}}{1992}]{Tomisaka92}
{Tomisaka} K.,  1992, \pasj, 44, 177

\bibitem[\protect\citeauthoryear{{Tomisaka}}{{Tomisaka}}{1998}]{Tomisaka98}
{Tomisaka} K.,  1998, \mnras, 298, 797

\bibitem[\protect\citeauthoryear{{Tomisaka} \& {Ikeuchi}}{{Tomisaka} \&
  {Ikeuchi}}{1986}]{Tomisaka86}
{Tomisaka} K.,  {Ikeuchi} S.,  1986, \pasj, 38, 697

\bibitem[\protect\citeauthoryear{{Vogel}, {Kulkarni} \&
  {Scoville}}{\protect\mniiiauthor{Vogel88}{{Vogel}, {Kulkarni} \&
  {Scoville}}{{Vogel} et~al.}}{1988}]{Vogel88}
{Vogel} S.~N.,  {Kulkarni} S.~R.,    {Scoville} N.~Z.,  1988, \nat, 334, 402

\bibitem[\protect\citeauthoryear{{Weaver}, {McCray}, {Castor}, {Shapiro} \&
  {Moore}}{{Weaver} et~al.}{1977}]{Weaver77}
{Weaver} R.,  {McCray} R.,  {Castor} J.,  {Shapiro} P.,    {Moore} R.,  1977,
  \apj, 218, 377

\bibitem[\protect\citeauthoryear{{Welsh}, {Sallmen} \&
  {Jelinsky}}{\protect\mniiiauthor{Welsh05}{{Welsh}, {Sallmen} \&
  {Jelinsky}}{{Welsh} et~al.}}{2005}]{Welsh05}
{Welsh} B.~Y.,  {Sallmen} S.,    {Jelinsky} S.,  2005, \aap, 440, 547

\bibitem[\protect\citeauthoryear{{West}, {English}, {Normandeau} \&
  {Landecker}}{{West} et~al.}{2007}]{West07}
{West} J.~L.,  {English} J.,  {Normandeau} M.,    {Landecker} T.~L.,  2007,
  \apj, 656, 914

\bibitem[\protect\citeauthoryear{{Williamson}, {Sanders}, {Kraushaar},
  {McCammon}, {Borken} \& {Bunner}}{{Williamson} et~al.}{1974}]{Williamson74}
{Williamson} F.~O.,  {Sanders} W.~T.,  {Kraushaar} W.~L.,  {McCammon} D.,
  {Borken} R.,    {Bunner} A.~N.,  1974, \apjl, 193, L133

\bibitem[\protect\citeauthoryear{{Zinnecker} \& {Yorke}}{{Zinnecker} \&
  {Yorke}}{2007}]{Zinnecker07}
{Zinnecker} H.,  {Yorke} H.~W.,  2007, \araa, 45, 481

\bibitem[\protect\citeauthoryear{{de Avillez} \& {Breitschwerdt}}{{de Avillez}
  \& {Breitschwerdt}}{2005}]{deAvillez05}
{de Avillez} M.~A.,  {Breitschwerdt} D.,  2005, \aap, 436, 585

\bibitem[\protect\citeauthoryear{{van Leeuwen}}{{van
  Leeuwen}}{2007}]{Vanleeuwen07}
{van Leeuwen} F.,  2007, \aap, 474, 653

\end{thebibliography}

\appendix
	
\section{PHYSICAL PROPERTIES OF THE BUBBLE}
\label{properties}

	For a Kompaneets model, the interior pressure, age, interior temperature, expansion velocity, and wall density of the superbubble can be determined by specifying the driving luminosity and the initial density of the exponential atmosphere at the height of the source. The superbubble wall density calculation is the same as done in \citet{Pon14a} for the Eridanus filament densities and thus is not reproduced here. \citet{Pon14a} find that the density of the wall should be between 1 and 6 cm$^{-3}$. A typically used value for the average rate of mechanical energy deposited into the ISM by an OB association is 10$^{38}$ erg s$^{-1}$ (e.g. \citealt{Heiles87, MacLow88,Shull95,Bally01}), although the Orion star-forming region is believed to have an average mechanical energy input rate of only $2 \times 10^{37}$ erg s$^{-1}$ \citep{Reynolds79, Brown94}. We choose to adopt an initial density range of 0.5-1 cm$^{-3}$ for the exponential atmosphere \citep{Heiles76, Ferriere91, Brown95, Kalberla09}, which is the same range chosen for the ionization front fitting in Section \ref{ion front}. 

\subsection{Pressure}
\label{pressure}
	
	X-ray data have been used to determine that the interior pressure, $P / k$, of the superbubble is $1 \times 10^4$-$5 \times 10^4$ cm$^{-3}$ K \citep{Burrows93, Guo95, Burrows96}. \citet{Burrows96}, however, suggest that the pressure, $P / k$, at the Eridanus end of the bubble might be as low as 1200 cm$^{-3}$ K.
		
	One of the known problems with a Kompaneets model is that it predicts that the top of an expanding superbubble will reach an infinite distance from the driving source in a finite time. Realistic bubbles will not obtain the high, late-time velocities predicted by a Kompaneets model. At later times in a bubble's evolution, when the expansion velocity given by the Hugoniot jump conditions becomes larger than the sound speed inside of the bubble, a pressure gradient is likely to form with the pressure being greatest towards the base of the atmosphere. At this time, the Kompaneets model assumption of constant pressure becomes invalid. Deviations away from exponential atmosphere profiles at large heights may also limit the expansion velocities of real superbubbles.   
	
	Because a Kompaneets model overpredicts the late-stage expansion velocity of a superbubble, such a model is also expected to slightly underpredict the pressure at late times of a bubble's expansion. For our adopted range of initial densities of the exponential atmosphere, the thermal pressure, $P / k$, within the bubble is between 0.9 and $1.1 \times 10^4$ K cm$^{-3}$ in model A. \citet{Basu99} suggest that for a bubble fit with a Kompaneets model, the best estimate for the pressure within the bubble at later times in the bubble's evolution is given by the interior pressure of the Kompaneets model when the expansion speed of the top of the bubble reaches the sound speed within the bubble. They argue that when the expansion speed becomes supersonic, a pressure gradient will form within the bubble and the interior pressure of the bottom section of the bubble will remain roughly constant throughout the remainder of the bubble's evolution. Using this approach, we find that a more accurate estimate of the pressure, $P / k$, for model A is $1.5 \times 10^4$ K cm$^{-3}$. This pressure is consistent with observationally derived pressures for the superbubble. For reference, the top cap of the superbubble becomes supersonic when $\tilde y$ is 1.886 in model A. The typical ISM pressure, $P / k$, is only on the order of $3 \times 10^3$ K cm$^{-3}$ (e.g. \citealt{Kalberla09}), and thus, the superbubble is clearly overpressured and should be expanding, as observed (e.g. \citealt{Reynolds79}).
	
\subsection{Age}
\label{age}
Because Kompaneets models tend to overpredict the expansion velocities of superbubbles, they are known to also slightly underpredict the ages of superbubbles \citep{Komljenovic99, Stil09}. Thus, the ages derived below for model A should be considered lower limits. \citet{Stil09} fit Kompaneets models to simulations of superbubbles expanding into stratified, magnetic media and show that while their Kompaneets model ages are too small, they are still accurate to within a factor of 4. \citet{Stil09}, however, do not simulate bubbles as highly elongated as the Orion--Eridanus superbubble. 

Model A predicts an age of approximately 2.5 Myr for the Orion--Eridanus superbubble. This age is consistent with previously calculated dynamical ages of the superbubble, which are on the order of a few Myr (e.g. \citealt{Brown94}). It is also in agreement, given that it may be underestimated by a factor of 4, with the ages of Orion OB1b, c, and d, 2-8, 2-6, and $<$2 Myr, respectively \citep{Brown94, Bally08}, but does not provide any significant constraints on which group initially formed the superbubble. Orion OB1a, with an age of 8-12 Myr \citep{Brown94}, may be marginally too old to be the driving source of the superbubble. 

\subsection{Interior Bubble Temperature}
\label{temperature}
Following the methodology of \citet{Basu99}, we assume that the interior temperature of the superbubble follows the equation from \citet{Weaver77} for the temperature at the centre of a spherical bubble:
\begin{equation}
T = 63.6 \left(\frac{L_0}{R_s}\right) \mbox{K},
\end{equation}
where $T$ is the bubble temperature, $L_0$ is the mechanical energy luminosity of the bubble, and $R_s$ is the equivalent spherical radius given by
\begin{equation}
R_s = \left(\frac{125}{154\pi}\right)^{1/5} L_0^{1/5} \rho_0^{-1/5} t^{3/5},
\end{equation}
where $\rho_0$ is the initial density at the height of the driving source and $t$ is the time since the formation of the superbubble. From this equation, model A should have an interior temperature of $4 \times 10^6$ K.  The photoevaporation of any small gas clumps remaining within the superbubble wall will cool the bubble interior and this quenching may cause the temperature of the interior to be slightly overestimated \citep{McKee84}. 

From X-ray observations, the interior temperature of the superbubble has been observationally determined to be 1-4 $\times 10^6$ K \citep{Williamson74, Naranan76, Long77, Burrows93, Guo95}, which is consistent with our model temperature. The interior density of the superbubble has also been estimated to be between 0.004 and 0.015 cm$^{-3}$, roughly consistent with the estimated temperature and pressure of the interior \citep{Burrows93, Guo95, Burrows96}.

\subsection{Expansion Speed}
\label{speed}

In model A, the expansion velocity of the bubble at the location of Arc A is roughly 120 km s$^{-1}$. Given that the angle between the expansion velocity of Arc A and the line of sight to Arc A in model A is small, 7$^{\circ}$, the expected observed line of sight velocity from this model, 119 km s$^{-1}$, is much higher than the observationally determined line of sight expansion velocity of 15 km s$^{-1}$ (e.g. \citealt{Reynolds79}). As discussed previously, it is a known problem of the Kompaneets model that the expansion velocities are overestimated at late times, particularly after the expansion speed of the top of the superbubble becomes supersonic. This transition to supersonic velocities occurs approximately 0.25 Myr before the bubble reaches its current configuration in model A. 

The expansion velocity of the bubble along its major axis must have been larger in the past than the currently observed line of sight expansion velocity, 15 km s$^{-1}$, since an average speed of 35 km s$^{-1}$ is required for the bubble to expand to its 300 pc total length, assuming an upper limit of 8 Myr for the age of the bubble, based upon the time since the formation of Orion OB1b \citep{Brown94}. This average speed of 35 km s$^{-1}$ is still small compared to the model prediction for the current expansion speed of the superbubble. If the bubble is only 2 Myr old, the required average expansion speed would be even larger, 140 km s$^{-1}$. 

\end{document}